\documentclass[reprint,amsmath,amssymb,aps,superscriptaddress,pre]{revtex4-2}
\usepackage[english]{babel}
\usepackage{amsmath,amssymb}
\usepackage{mathtools}
\usepackage{placeins}

\usepackage[utf8]{inputenc}
\usepackage{amsmath}
\usepackage{amsfonts}
\usepackage{amssymb}
\usepackage{amsthm}
\usepackage{amsopn}
\usepackage{upgreek}
\usepackage{bm}
\usepackage{epsfig}
\usepackage{relsize}
\usepackage{color}
\usepackage{hyperref}
\usepackage{verbatim}
\usepackage{siunitx}
\usepackage{nicefrac}
\usepackage{afterpage}
\usepackage{array}
\usepackage{courier}
\usepackage{booktabs}
\usepackage{longtable}

\usepackage[T1]{fontenc}

\newcommand{\rhoN}{\rho_{\text{N}}}
\newcommand{\rhoP}{\rho_{\text{P}}}
\newcommand{\rhoS}{\rho_{\text{S}}}
\newcommand{\rhoU}{\rho_{\text{U}}}
\newcommand{\rhoNb}{\rho_{\text{Nb}}}
\newcommand{\rhoNo}{\rho_{\text{N1}}}

\newcommand{\ZN}{Z_{\text{N}}}

\newcommand{\vN}{v_{\text{N}}}
\newcommand{\vP}{v_{\text{P}}}
\newcommand{\NP}{N_{\text{P}}}

\begin{document}

\title{Cooperation and competition of basepairing and electrostatic interactions in mixtures of DNA nanostars and polylysine}

\author{Gabrielle R. Abraham}
\altaffiliation{Equal contribution. Current address: Physics Department, University of Southern California}
\affiliation{Physics Department, University of California, Santa Barbara}
\author{Tianhao Li}
\altaffiliation{Equal contribution. Current address: Simons Center for Computational Physical Chemistry, Department of Chemistry, New York University, New York}
\affiliation{Chemistry Department, Princeton University}
\author{Anna Nguyen}
\affiliation{Biomolecular Science and Engineering Program, University of California, Santa Barbara}
\author{William M. Jacobs}
\email{wjacobs@princeton.edu}
\affiliation{Chemistry Department, Princeton University}
\author{Omar A. Saleh}
\email{saleh@ucsb.edu}
\affiliation{Physics Department, University of California, Santa Barbara}
\affiliation{Biomolecular Science and Engineering Program, University of California, Santa Barbara}
\affiliation{Materials Department, University of California, Santa Barbara}

\begin{abstract}
  Phase separation in biomolecular mixtures can result from multiple physical interactions, which may act either complementarily or antagonistically.
  In the case of protein--nucleic acid mixtures, charge plays a key role but can have contrasting effects on phase behavior.
  Attractive electrostatic interactions between oppositely charged macromolecules are screened by added salt, reducing the driving force for coacervation.
  By contrast, base pairing interactions between nucleic acids are diminished by charge repulsion and thus enhanced by added salt, promoting associative phase separation.
  To explore this interplay, we combine experiment and theory to map the complex phase behavior of a model solution of poly-L-lysine (PLL) and self-complementary DNA nanostars (NS) as a function of temperature, ionic strength, and macromolecular composition.
  Despite having opposite salt dependences, we find that electrostatics and base pairing cooperate to stabilize NS--PLL coacervation at high ionic strengths and temperatures, leading to two- or three-phase coexistence under various conditions.
  We further observe a variety of kinetic pathways to phase separation at different salt concentrations, resulting in the formation of nonequilibrium aggregates or droplets whose compositions evolve on long timescales.
  Finally, we show that the cooperativity between electrostatics and base pairing can be used to create immiscible coacervates that partition various NS species at intermediate salt concentrations.
  Our results illustrate how the interplay between distinct interaction modes can greatly increase the complexity of the phase behavior relative to systems with a single type of interaction.
\end{abstract}

\maketitle

\section{Introduction}
Solutions of macromolecules in solvent can spontaneously separate into spatially distinct regions that are, respectively, dense and dilute in the macromolecules. 
Frequently, the dense phase retains liquid-like properties, and the process is termed liquid-liquid phase separation (LLPS). 
LLPS is driven by net attractive interactions between macromolecules that are mediated by the solvent or cosolutes, such as electrostatic, hydrogen-bonding, or hydrophobic interactions.
These interactions are free energies, and thus account for both entropic and enthalpic contributions to the net thermodynamic driving force.
LLPS can be driven by interactions between molecules of the same species (``homotypic''), between molecules of different species (``heterotypic''), or a combination of the two.
    
Recently, interest in biomolecular LLPS has surged due to its role in the formation of biomolecular condensates, which provide spatial and temporal organization within living cells.
\textit{In vivo} LLPS of proteins and nucleic acids has been implicated in numerous processes, such as transcription and stress response~\cite{Sabari2020, Campos-Melo2021, Peng2020}.
Engineering biomolecular LLPS for technological applications is also of interest, for example to control cellular behavior for synthetic biology purposes~\cite{WAN2024108452} or to create multiphase \emph{in vitro} structures that can be used to build complex materials~\cite{song2024liquid} or bioreactors~\cite{LIM2024496}.

Predicting and controlling biomolecular LLPS in the multicomponent settings that are biologically and/or technologically relevant remains challenging due to the variety of entropic and energetic effects that can contribute to both homotypic and heterotypic interactions between various molecules.
Numerous studies have analyzed the effects   of specific interaction modalities---including  interactions between folded protein domains~\cite{li2012phase}, hydrophobic interactions between aromatic residues in prion-like domains in disordered proteins~\cite{martin2020valence}, electrostatic interactions in binary mixtures of oppositely charged disordered proteins~\cite{chowdhury2023driving}, and base pairing among RNA oligonucleotides~\cite{jain2017rna}---within controlled, \emph{in vitro} reconstitutions of biomolecular mixtures.
Nonetheless, further work is needed to explore the interplay among these various types of interactions, in particular in biomolecular mixtures exhibiting both homotypic and heterotypic interactions.

Here, we carry out an experimental and theoretical analysis of LLPS in the presence of both homotypic, base pairing driven association and heterotypic, electrostatically driven coacervation in a multicomponent biomolecular mixture with well-defined interactions.
Our experimental system is based on a model system of DNA nanoparticles termed nanostars (NSs), branched DNA particles that interact through single-stranded sequences (``sticky ends'')~\cite{Biffi2013}.
NSs can be designed with sticky ends that are palindromic, allowing homotypic binding with a neighboring  NS with an identical sticky end sequence and leading to associative LLPS.
Previous studies have investigated the effect on LLPS of various NS design parameters, including sequence-dependent intermolecular interactions~\cite{Jeon2018,Sato2020}, particle flexibility~\cite{Nguyen2017, Jeon2018, Lee2021}, size\cite{agarwal2022growth}, and valence~\cite{conrad2022emulsion, Conrad2019, Biffi2013, Bianchi2006}.
In this work, we explore the interplay between homotypic base pairing and heterotypic interactions due to electrostatic attraction between NSs and poly-L-lysine (PLL), a positively charged polypeptide that can form heterotypic droplets or gels when mixed with single- or double-stranded nucleic acids.
Previous studies have shown that the presence and nature of the condensed phase are sensitive to the salt conditions, as well as the structure of the nucleic acid, due to the dominant role of electrostatics in driving phase separation~\cite{Wee2021, Yin2016, Vieregg2018, wadsworth2022rnas, jain2017rna}.

The central aim of this work is to understand how the interplay between electrostatic interactions and  base pairing determines the phase behavior of NS+PLL mixtures.
In particular, we investigate the effects of charge screening by added salt on these two interaction modalities, which is well-known to strengthen base pairing interactions but weaken the driving force for electrostatic coacervation.
Using fluorescent microscopy, we observe a diversity of liquid- and gel-like phase behaviors of NS+PLL mixtures and, by characterizing each phase's molecular composition and temperature response, we determine the dominant interaction mode for the phase in a given set of conditions.
In parallel, we develop an analytical model of the free energy of the NS+PLL mixture that treats base pairing and electrostatics on equal footing and thus allows us to predict the phase behavior of the solution.
Experiments and theory both reveal rich phase behavior, including transitions between dense phases that are alternatively electrostatically or base pairing-dominated.
We further show that these two interaction modalities can either cooperate to stabilize a single dense phase or compete to form distinct dense phases.
Our findings indicate that the strengths of NS--NS and NS--PLL interactions are closely tuned in the studied conditions, permitting distinct phase behaviors to appear upon relatively modest environmental changes. 
Overall, our results demonstrate that rich phase behavior can arise from a simple set of molecular interactions---within a biologically relevant range of solution conditions---and provide a quantitative understanding of this diversity.

\begin{figure*}
  \centering
  \includegraphics[width=\textwidth]{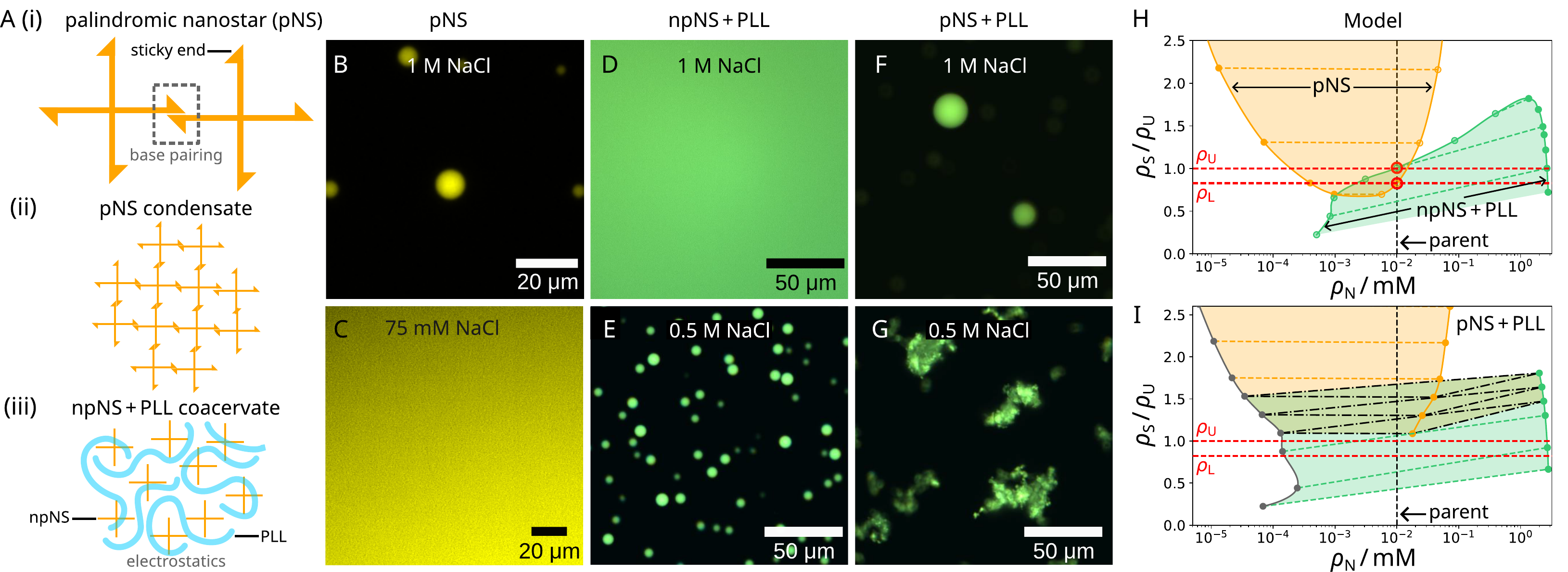}
  \caption{
    Overview of nanostar (NS) and poly-L-lysine (PLL) phase behavior.
    (A)~Schematics of (i) a four-armed DNA nanostar with palindromic sticky ends (pNS), (ii) associative phase separation driven by sticky-end base pairing, and (iii) coacervation of nonpalindromic nanostars (npNS) with PLL due to electrostatic interactions.
    (B-G) Fluorescent images of pNS (yellow), npNS (yellow), and/or PLL (cyan) mixtures, performed at \SI{10}{\micro M} NS and \SI{19.2}{\micro M} PLL. 
    (B)~A pNS solution phase separates at \SI{1}{M} NaCl, but (C)~remains in a single phase at \SI{0.075}{M} NaCl.
    (D)~A npNS+PLL solution exists as a single phase at \SI{1}{M} NaCl, but (E)~forms spherical, liquid-like droplets at \SI{0.5}{M} NaCl.
    (F)~By contrast, a pNS+PLL solution phase separates at \SI{1}{M} NaCl to form droplets, but (G)~is arrested in a gel-like aggregate at \SI{0.5}{M} NaCl.
    (H)~Overlay of two separate coexistence regions predicted from our theoretical model for solutions of, respectively, pNS (yellow) and npNS+PLL (green) as a function of NS concentration, $\rho_{\text{N}}$, and salt concentration, $\rho_{\text{S}}$.  In the npNS+PLL case, the solution is assumed to be macromolecularly charge balanced, so that $\rho_{\text{P}} = (192/100) \rho_{\text{N}}$.  Solid curves and points indicate coexistence curves, and dashed lines indicate representative tie lines.  The transition salt concentrations $\rho_{\text{U}}$ and $\rho_{\text{L}}$ are defined as the highest and lowest salt concentrations at which phase separation occurs in the npNS+PLL and pNS solutions, respectively, at the experimental parent concentration (vertical line, $\rho_{\text{N}} = \SI{10}{\micro M}$). The existence of $\rho_{\text{U}}$ and $\rho_{\text{L}}$, and the prediction that $\rho_{\text{L}}<\rho_{\text{U}}$, are consistent with experiment (panels B-E).
    (I)~Predicted phase diagram for macromolecularly charge-balanced pNS+PLL mixtures showing both two-phase (colored tie lines) and three-phase (black tie lines) coexistence regions. Two-phase regions are colored according to the composition of the condensed phase, $\rhoN / \rhoP \gg 1$ (yellow) or $\rhoN / \rhoP \sim 1$ (green).  The prediction of stable pNS+PLL coacervates at intermediate salt concentrations, $\rhoS \gtrsim \rho_{\text{U}}$, is consistent with experiment (panel F).  The predicted three-phase behavior at intermediate salt concentrations is also observed experimentally (see Fig.~\ref{fig:PD}).   Note that the tie lines are shown as straight lines for clarity but are in fact curved on these semilog phase diagrams.
    }
\label{fig:intro}
\end{figure*}

\section{Results}

\subsection{NS design and NS-only associative phase separation}
Experiments were carried out with DNA NSs that were self-assembled from four DNA oligomers by thermal annealing, following prior work~\cite{conrad2022emulsion} (Fig.~\ref{fig:intro}A).
Each NS consists of four double-stranded DNA arms that meet at a junction and terminate at the distal end in a single-stranded segment.
Each arm either terminates in an overhanging palindromic sticky-end sequence, 5'-{\fontfamily{qcr}\selectfont {\textit{CGATCG}}}-3', that is capable of binding a second identical segment, or a non-palindromic sequence that cannot engage in such homotypic interactions.
Fluorescent imaging revealed that solutions of palindromic NSs (pNSs) phase separate at high salt (Fig.~\ref{fig:intro}A,B and Fig.~\ref{fig:SI-NSalone}), whereas those with non-palindromic NSs (npNSs) were not found to phase separate under any conditions (Fig.~\ref{fig:SI-NSalone}).
Varying the salt concentration further showed that solutions with \SI{10}{\micro M} pNS do not phase separate at and below \SI{75}{mM} NaCl (Fig.~\ref{fig:intro}C and Fig.~\ref{fig:SI-NSalone}).
These observations are consistent with associative phase separation driven by the hybridization of pNS sticky ends above a lower critical salt concentration.
Decreasing the NaCl concentration reduces the driving force for phase separation, since the attractive base pairing interactions weaken due to reduced electrostatic screening between the charged sticky ends~\cite{santalucia2004thermodynamics}.

\subsection{Complex coacervation of NS and PLL}
To investigate the effects of electrostatically driven heterotypic interactions, we mixed npNSs with 100-residue long PLL in a concentration ratio chosen to match the macromolecular charge densities (i.e., \SI{10}{\micro M} npNS and \SI{19.2}{\micro M} PLL, assuming total charges of $-192e$ and $100e$ per molecule, respectively).
Both NSs and PLL were labeled with distinct fluorescent dyes to permit identification of the presence and composition of dense phases through fluorescent imaging.
We found that npNS+PLL mixtures form gels at low salt, coexisting dilute and liquid-like dense phases at intermediate salt, and a single mixed phase at high salt (Fig.~\ref{fig:intro}A,D,E and Fig.~\ref{fig:SI-npPhaseDiagram}).
The low-salt gel state is characteristic of kinetically-arrested phase separation, where material relaxation and densification is impeded by strong binding interactions between the components, while the appearance of liquid-like spherical droplets at intermediate salt concentrations indicates that the system has reached a local equilibrium.
These states are not dependent on the presence of non-palindromic sticky ends, as we observed similar behavior in mixtures of PLL and ``blunt'' NSs which lack sticky ends (Fig.~\ref{fig:SI-npPhaseDiagram}).
These results are consistent with prior nucleic-acid/PLL studies~\cite{Vieregg2018}, indicating that npNS+PLL dense phases are predominantly stabilized by heterotypic electrostatic attractions.
This behavior, commonly referred to as (complex) coacervation, exhibits an upper critical salt concentration, since higher salt concentrations more effectively screen the heterotypic electrostatic attractions and also reduce the entropy gain from counterion release~\cite{sing2020recent} (although entropic effects of the solvent also play an important role \cite{priftis2012thermodynamic,chen2022driving}).
Conversely, sufficiently low salt concentrations strengthen these effects, and so oppose relaxation of the dense phase~\cite{Vieregg2018,wang2014polyelectrolyte,jawerth2018salt,yuan2025network}, consistent with our observations of gel structures at \SI{0.25}{M} NaCl (Fig.~\ref{fig:SI-npPhaseDiagram}).

\subsection{Mixed interaction modalities}
Having established that the individual homotypic and heterotypic interactions respond oppositely to changes in salt concentration due to the respective lower and upper critical behavior of pNS and npNS+PLL solutions, we next investigated the phase behavior with both types of interactions present.
We created macromolecularly charge-balanced pNS+PLL mixtures (\SI{10}{\micro M} pNS and \SI{19.2}{\micro M} PLL) at various salt concentrations.
These mixtures exhibited both a low-salt gel phase and liquid-like phase coexistence at higher salt concentrations (Fig. \ref{fig:intro}F,G).
Yet compared to npNS+PLL mixtures, the pNS+PLL gel/liquid transition occurs at a higher salt concentration, indicating a broader regime of kinetically-trapped gel behavior and thus stronger net interactions among the components.
Moreover, liquid-like droplets persist to \SI{1}{M} NaCl (Fig.~\ref{fig:intro}F) and remain stable over several days (Fig.~\ref{fig:SI-Stability}), indicating that they have reached a local equilibrium.
These droplets are composed of both pNS and PLL, indicating that heterotypic electrostatic interactions contribute to the stabilization of the dense phase under conditions where electrostatic interactions alone do not lead to coacervation.
These observations provide evidence of cooperativity between the heterotypic and homotypic interactions, suggesting that the combination of electrostatics and base pairing increases the net strength of the attractive interactions among the NS and PLL molecules.

\subsection{Theoretical model development}
To facilitate the interpretation of these observations, we developed a theoretical model to investigate the equilibrium and kinetic behavior of NS+PLL mixtures.
This model combines a DLVO~\cite{gaussian-blob} description of screened electrostatics, which is relevant for the monovalent salts used in our experiments, with a statistical associating fluid theory (SAFT)~\cite{chapman1989saft} treatment of sticky-end hybridization that has previously been shown to accurately describe NS-only phase behavior~\cite{rovigatti2014accurate}.
The model accounts for both entropic and enthalpic contributions to base pairing and electrostatics, which, importantly, are considered on equal footing within a consistent mean-field framework (see Methods).
The model only further includes the entropies of translational degrees of freedom and excluded volume, and so is relatively simple and interpretable.
All model parameters are based on experimental measurements, including a temperature independent Bjerrum length for screened electrostatics that implicitly accounts for solvent effects~\cite{archer1990dielectric}, the temperature and salt-dependence of the sticky-end hybridization free energies~\cite{santalucia2004thermodynamics,locatelli2017condensation}, and the physical properties of the NS and PLL molecules.
The lack of free parameters in our model, which differs from other recent approaches to model coacervate systems~\cite{spruijt2010binodal,galvanetto2023extreme,lin2019narrow}, is a strength that allows us to make direct mechanistic interpretations.
Details of the model derivation and parametrization are provided in Sections~S1--S5.

Applying this model to the pNS-only and npNS+PLL systems produced equilibrium predictions consistent with the experimental observations and the expected salt dependence of the two separate interaction modalities.
In the case of the pNS-only mixture, the model predicts a lower critical salt concentration and a lower transition salt concentration, $\rho_{\text{L}}$, defined by the intersection of the parent concentration (\SI{10}{\micro M} pNS) and the binodal (Fig.~\ref{fig:intro}H).
By contrast, the npNS+PLL mixture exhibits an upper critical salt concentration and an upper transition salt concentration, $\rho_{\text{U}}$, at the parent concentration (\SI{10}{\micro M} npNS, \SI{19.2}{\micro M} PLL).
In Fig.~\ref{fig:intro}H, we show the binodal for a macromolecularly charge-balanced mixture, in which case the PLL:npNS concentration ratio is the same in the two phases by symmetry.
This binodal has the typical shape of DLVO-based theoretical predictions for symmetric polyelectrolyte solutions near the critical point~\cite{sing2017development}, although the predicted dilute-phase concentration at low salt is influenced by the charge and size asymmetry of the npNS and PLL molecules in our model.
The values of $\rho_{\text{U}} = \SI{0.23}{M}$ and $\rho_{\text{L}} = \SI{0.19}{M}$ each differ by roughly a factor of three relative to the experimental values of \SI{0.95}{M} and \SI{0.075}{M}, respectively; however, the crucial qualitative prediction that $\rho_{\text{U}} > \rho_{\text{L}}$ is consistent with the experimental observations.
This level of agreement is sufficient to distinguish the effects of cooperative versus competitive homotypic and heterotypic interactions by comparing theoretical predictions with experimental results.
Therefore, throughout the remainder of this article, we normalize our predicted salt concentrations relative to $\rho_{\text{U}}$ when making comparisons with experiments.

When applied to the pNS+PLL system, our model predicts that cooperative phase behavior is due to the comparable strengths of the two interaction modalities at intermediate salt.
In particular, for salt concentrations between the upper and lower critical salts of the npNS+PLL and pNS systems, respectively, the model predicts that homotypic pNS interactions substantially reduce the concentration of macromolecules in the dilute phase that coexists with the coacervate (Fig.~\ref{fig:intro}I).
This observation implies an increase in the stability of the condensed coacervate phase due to homotypic pNS interactions, an effect we confirmed by carrying out a `dominance analysis'~\cite{qian2024dominance} of the free-energy landscape predicted by our model (see Fig.~\ref{fig:SI-dominance} and Section~\ref{sec:SI-dominance}).
Overall, these predictions are consistent with the increased stability of coacervate droplets upon addition of homotypic pNS interactions (Fig.~\ref{fig:intro}F).
Further, for salt concentrations above approximately $\rho_{\text{U}}$, the model predicts equilibrium three-phase coexistence for a range of parent concentrations  (Fig.~\ref{fig:intro}I), indicating that coacervate droplets can stably coexist with both the dilute phase and majority-NS droplets at salt concentrations well above $\rho_{\text{U}}$.
Three coexisting phases are indeed experimentally observed (see below), though in a manner dependent on kinetic considerations.

\subsection{Temperature dependence of mixed-modality phase behavior}
To test the hypothesis that homotypic base pairing is essential for pNS+PLL phase separation at $1$~M salt concentrations, we examined the temperature response of pNS+PLL droplets.
Homotypic pNS droplets (with no PLL present) are known to exhibit upper critical solution temperatures (UCSTs) due to the melting of sticky-end duplexes at elevated temperatures~\cite{Biffi2013, conrad2022emulsion}.
By contrast, electrostatically stabilized droplets can exhibit more complicated responses to temperature, but frequently exhibit lower critical solution temperatures (LCSTs) due to the dominant entropic contributions to coacervation~\cite{ali2019lower}.
We therefore sought to distinguish homotypic and heterotypic interactions via their temperature dependence.

\begin{figure}
    \centering
    \includegraphics[width=\columnwidth]{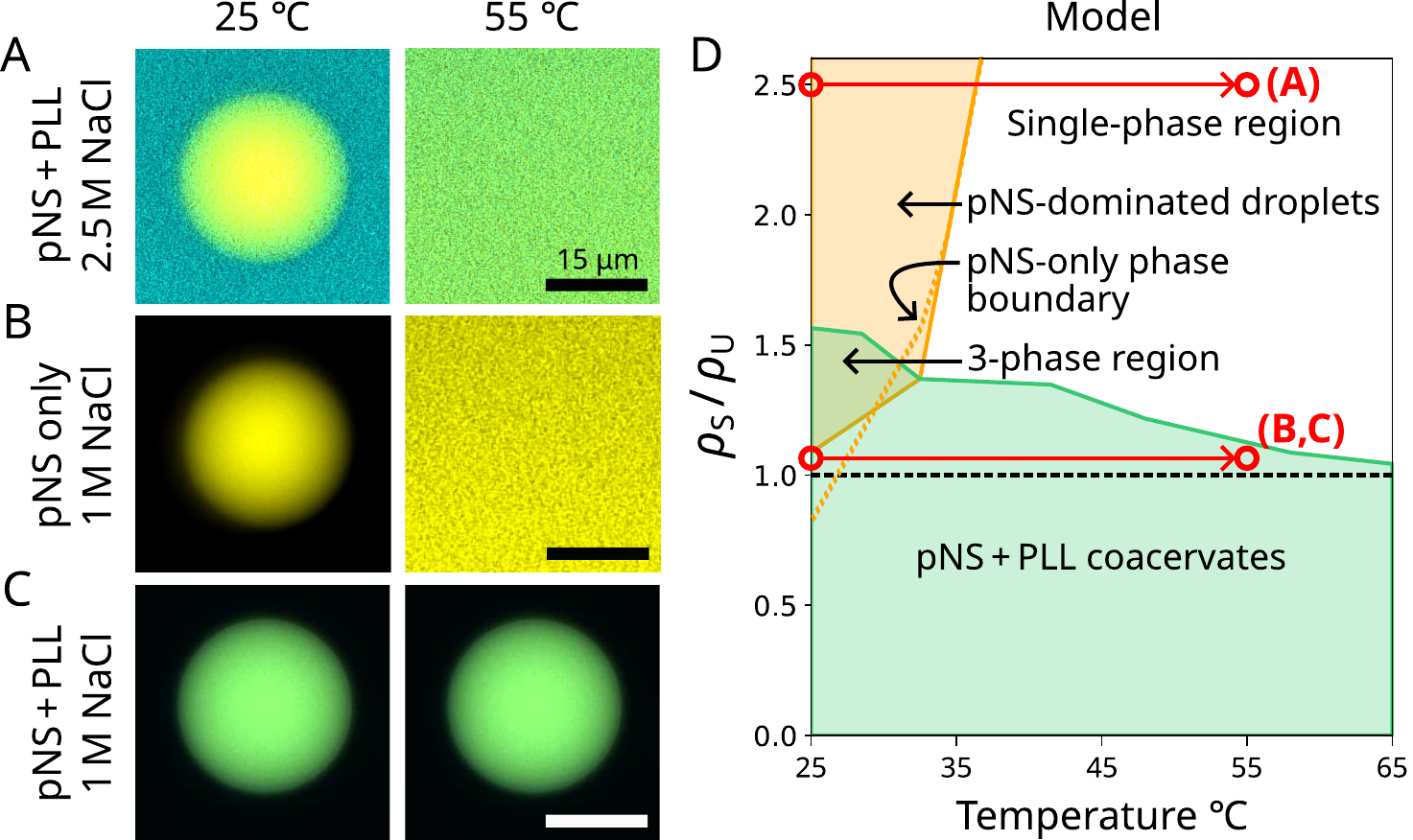}
    \caption{Temperature dependence of single and mixed-modality phase behavior with pNS (yellow) and PLL (cyan).
      (A)~At \SI{2.5}{M} NaCl, pNS+PLL solutions form NS-dominated liquid-like droplets that are stable at \SI{25}{\celsius} but melt at \SI{55}{\celsius}.
      (B)~In the absence of PLL, pNS droplets similarly melt at \SI{55}{\celsius} at the lower salt concentration of at \SI{1}{M} NaCl.
      (C)~By contrast, pNS+PLL solutions at \SI{1}{M} NaCl form liquid-like droplets that are stable at both \SI{25}{\celsius} and \SI{55}{\celsius}.
      (D)~Predicted pNS+PLL phase diagram as a function of temperature and salt concentration at the parent concentration of \SI{10}{\micro M} pNS and \SI{19.2}{\micro M} PLL, colored according to the composition of the condensed phase (as in Fig.~\ref{fig:intro}I).  The phase boundary for a \SI{10}{\micro M} pNS-only solution is indicated by a dashed yellow line.  The red arrows correspond to the heating experiments, and agree with the observed UCST behavior at high salt concentrations (panel A) and the lack of melting at intermediate salt concentrations (panel C).  These observations indicate that the pNS-dominated droplets are stabilized primarily by base pairing interactions, whereas the pNS+PLL coacervates are stabilized by a combination of electrostatic and base pairing interactions, in agreement with a `dominance analysis' \cite{qian2024dominance} of the predicted free-energy landscape (see Fig.~\ref{fig:SI-dominance} and Section \ref{sec:SI-dominance}).
  }
    \label{fig:temp}
\end{figure}

We first verified the expected UCST behavior of homotypically stabilized droplets.
At a high salt concentration of \SI{2.5}{M} NaCl, where increased screening results in negligible electrostatic interactions and enhanced base pairing, droplets in a pNS+PLL mixture melt upon increasing the temperature from \SI{25}{\celsius} to \SI{55}{\celsius} (Fig.~\ref{fig:temp}A).
This behavior is analogous to that of pNS-only solutions in which droplets undergo a melting transition between the same two temperatures at \SI{10}{\micro M} pNS and \SI{1}{M} NaCl (Fig.~\ref{fig:temp}B).
As anticipated, these observations are consistent with base pairing-driven phase separation.

If pNS+PLL phase separation were dominated by homotypic base pairing at the intermediate salt concentration of \SI{1}{M} NaCl, then we would expect the pNS+PLL droplets to also dissolve at high temperatures.
However, pNS+PLL coacervates are stable for many hours at \SI{55}{\celsius} at \SI{1}{M} NaCl (Fig.~\ref{fig:temp}C).
We therefore conclude that because neither electrostatic coacervation nor base pairing alone is sufficient to stabilize droplets under these conditions, phase separation must occur as a result of cooperative heterotypic and homotypic interactions, as also predicted by our model.

These observations also agree qualitatively with the temperature dependence predicted by our model.
In the model, the temperature dependence of base pairing is introduced through empirical, sequence-dependent sticky-end hybridization free energies~\cite{santalucia2004thermodynamics}.
By contrast, the electrostatic contribution to the mean-field free energy is independent of temperature due to the assumption of a constant Bjerrum length~\cite{archer1990dielectric}.
The resulting temperature and salt-dependent predictions for a solution with a parent concentration of \SI{10}{\micro M} pNS and \SI{19.2}{\micro M} PLL are shown Fig.~\ref{fig:temp}D.
At high salt ($\rhoS \gg \rhoU$), the predicted phase boundary for pNS-dominated droplets closely follows the phase boundary for pNS-only solutions at the parent concentration of \SI{10}{\micro M} pNS, since base pairing is the dominant driving force for phase separation in both of these cases.
However, in pNS+PLL solutions at intermediate salt ($\rhoS \gtrsim \rhoU$), cooperativity among the two driving forces shifts the phase boundary for coacervation, which occurs at $\rhoU$ in the absence of base pairing interactions by definition, to higher salt concentrations  (Fig.~\ref{fig:SI-shift}).
This predicted increase in the range of conditions at which coacervation can occur is consistent with the observation of pNS+PLL phase separation at intermediate salt concentrations and high temperatures in Fig.~\ref{fig:temp}C.

\subsection{Nonequilibrium phase behavior revealed by varying macromolecular composition}
We sought to distinguish the contributions of homotypic and heterotypic interactions by varying the macromolecular composition.
Electrostatically driven coacervation is typically promoted by macromolecular charge ratios close to unity~\cite{Vieregg2018, zhang2005phase}, leading to phase separation in which both the dilute and condensed phases are macromolecularly charge balanced.
By contrast, phase separation due to homotypic base pairing interactions requires no neutralizing polycation, and so is likely less sensitive to the macromolecular charge ratio.
Therefore, to investigate the role of charge stoichiometry, we explored the phase behavior at \SI{25}{\celsius} by holding the pNS concentration fixed at $\rhoN = \SI{10}{\micro M}$ and tuning the PLL concentration $\rhoP$.

\begin{figure*}
 \centering
 \includegraphics[width=\textwidth]{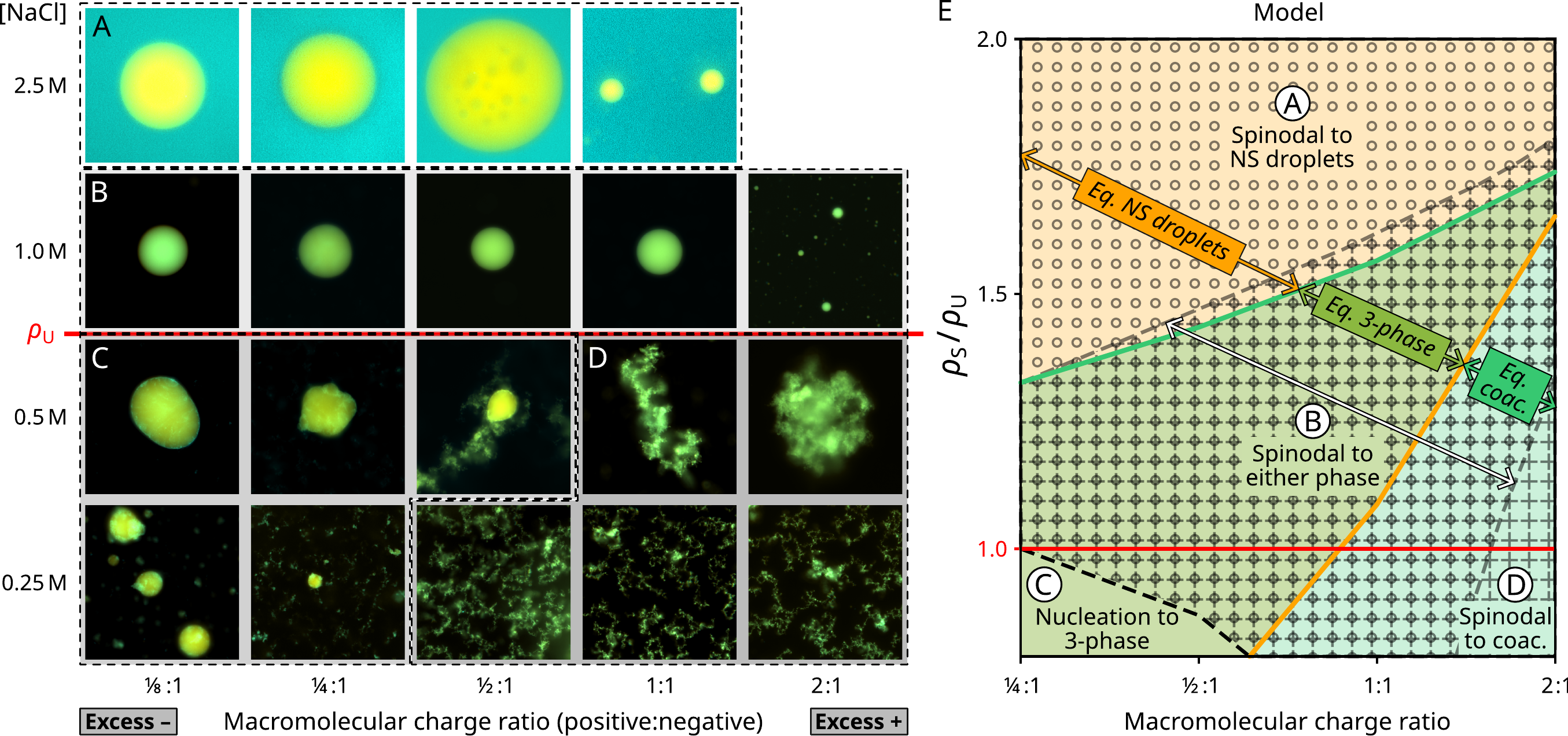}
 \caption{
   (A--D) Fluorescent images of condensed phases formed in pNS+PLL mixtures (pNS: yellow; PLL: cyan) across various compositions and salt concentrations, and at constant pNS concentration, $\rhoN = \SI{10}{\micro M}$, at \SI{25}{\celsius}.
   (A)~At \SI{2.5}{M}, pNSs form liquid-like droplets that exclude PLL immediately after mixing, but permit PLL partitioning over time scales of days (Fig~\ref{fig:SI-slowDynamics}).
   (B)~At \SI{1}{M} NaCl, liquid-like droplets form, enriched in both pNSs and PLL, and are stable over days (Fig.~\ref{fig:SI-Stability}A).
   (C)~At lower salt concentration, and when pNS is present in sufficient excess, three-phase coexistence is observed, with pNS-enriched liquid-like droplets coexisting with, but remaining spatially distinct from, pNS+PLL gels; this morphology is  stable for over 48 hours (Fig.~\ref{fig:SI-Stability}B).
   (D)~At lower salt concentration, but higher PLL concentrations, pNSs and PLL form a single gel phase that is stable for over 48 hours (Fig.~\ref{fig:SI-Stability}C).
   (E)~Predicted phase diagram showing both equilibrium and kinetic boundaries for the conditions measured in panels A--D. Equilibrium behavior is demarcated by solid boundaries, and displays two-phase coexistence, with either pNS-enriched droplets (yellow region) or pNS+PLL coacervates (cyan region), or three-phase coexistence involving both dense phases (green region), depending on the salt concentration, $\rhoS$, and the PLL concentration, $\rhoP = (192/100)\times(\text{charge ratio})\times\SI{10}{\micro M}$. Dashed lines delineate kinetic differences, indicating where the solution is predicted to undergo spinodal decomposition (to pNS-enriched droplets, open circles, and/or pNS+PLL coacervates, crosses), or, alternatively, nucleation from a metastable parent phase (no pattern).  The circled labels emphasize the qualitative agreement between the predicted behaviors and the experiments, including (A) rapid pNS phase separation leading to pNS-enriched droplets, (B) rapid phase separation leading to pNS+PLL droplets at salt concentrations above $\rho_{\text{U}}$, (C) three-phase behavior accessed via slower nucleation at salt concentrations below $\rho_{\text{U}}$ and low macromolecular charge ratio, and (D) the existence of a single condensed coacervate phase at low salt concentrations and high macromolecular charge ratio.}
\label{fig:PD}
\end{figure*}
  
We found that phase separation is insensitive to the macromolecular charge ratio for salt concentrations at and above \SI{1}{M} NaCl, consistent with the expectation when base pairing significantly stabilizes the dense phase (Fig.~\ref{fig:PD}A,B).
At \SI{2.5}{M} NaCl, pNS-enriched droplets rapidly formed at four charge ratios ranging from \nicefrac{1}{8}:1 to 1:1 PLL:pNS (Fig.~\ref{fig:PD}A).
These droplets were found to exclude PLL shortly after mixing, but then displayed an increasing enrichment of PLL over several days (Fig.~\ref{fig:SI-slowDynamics}) that was accelerated at higher excess-negative macromolecular charge ratios and shorter PLL lengths (Fig.~\ref{fig:SI-slow-dynamicsPLL20}).
Nonetheless, homotypic interactions appear to provide the dominant initial driving force for phase separation under these conditions, and the pNS remain more strongly enriched in the droplets than the PLL after the long equilibration period (Fig.~\ref{fig:SI-2.5MPartitioning}).
At \SI{1}{M} NaCl, we similarly observed charge ratio-independent phase behavior for macromolecular charge ratios up to 2:1 PLL:pNS (Fig.~\ref{fig:PD}B).
Under these conditions, pNSs and PLL quickly colocalize in the dense phase after mixing, and the droplets persist for several days (Fig.~\ref{fig:SI-Stability}).
These observations are consistent with phase separation to a local equilibrium stabilized by cooperative homotypic and heterotypic interactions at \SI{1}{M} NaCl.

Varying the macromolecular charge ratio has a greater effect at lower salt concentrations, since the heterotypic electrostatic interactions play a more significant role under these conditions.
When the polyanion is in sufficient excess (below charge ratios of approximately \nicefrac{1}{2}:1 PLL:pNS) at salt concentrations below \SI{0.5}{M} NaCl, we observed coexistence of the dilute phase with two distinct dense phases: pNS-enriched droplets and pNS+PLL coacervate gels (Fig.~\ref{fig:PD}C).
Such three-phase coexistence is not surprising given that these salt conditions independently allow formation of both pNS droplets (in pNS-only solutions) and kinetically arrested coacervate gels (in macromolecularly charge-matched pNS+PLL solutions).
Frequently, we found that the gel forms a layer encasing a larger pNS-enriched droplet, suggesting that the coacervate gels have a lower interfacial tension with the dilute solution and thus preferentially coat the droplets.
Moreover, the direct contact between the droplets and coacervate gels under these conditions suggests that complete wetting is thermodynamically favored, which is likely due to the presence of pNS in both phases.
By contrast, the phase behavior of mixtures with excess polycation (2:1 PLL:pNS)  at or below \SI{0.5}{M} (Fig.~\ref{fig:PD}D) is similar to that of charge balanced systems (Fig.~\ref{fig:intro}G), showing  kinetically arrested gel morphologies.

Our theoretical model produces equilibrium and kinetic predictions (Fig.~\ref{fig:PD}E) that account for the distinct behaviors experimentally observed in Fig.~\ref{fig:PD}A--D, and the qualitative conditions at which they occur.
For the salt conditions explored (from $\rhoS \lesssim \rhoU$ to $\rhoS \gg \rhoU$), the model predicts equilibrium pNS-enriched droplets at high salt and excess polyanion (as in Fig.~\ref{fig:PD}A); NS+PLL coacervates at low salt and excess polycation (as in Fig.~\ref{fig:PD}D), and a three-phase region in which both condensed phases coexist at intermediate salt concentrations and macromolecular charge ratios (as in Fig.~\ref{fig:PD}C).
(Note that the equilibrium predictions at the 1:1 charge ratio in Fig.~\ref{fig:PD}E correspond to the \SI{25}{\celsius} phase behavior in Fig.~\ref{fig:temp}D and the $\rhoS$ versus $\rhoN$ projection of the phase diagram in Fig.~\ref{fig:intro}I.)
The overall topology of the phase diagram in Fig.~\ref{fig:PD}E, and in particular the diagonal phase boundary between three-phase coexistence and NS+PLL coacervates at low salt concentrations, agrees with the experimental observations.
The model also predicts that PLL is enriched in the condensed phase relative to the dilute phase at high salt concentrations ($\rhoS \gg \rhoU$) at equilibrium, although to a lesser extent than pNSs since phase separation is primarily driven by base pairing interactions under these conditions (Fig.~\ref{fig:SI-pc}).
This prediction is consistent with the long-time behavior of the pNS-dominated droplets at \SI{2.5}{M} NaCl, as noted above.

Explaining the precise experimental conditions under which three-phase coexistence is observed in pNS+PLL mixtures requires us to consider kinetic effects in addition to equilibrium phase behavior.
According to our model, the three-phase region seen in Fig.~\ref{fig:PD}C extends at equilibrium into the regions probed in Fig.~\ref{fig:PD}B and, to a lesser extent, Fig.~\ref{fig:PD}D.
However, the model also predicts that phase separation from an initially homogeneous solution proceeds via spinodal decomposition, since the minimum-free-energy path from the parent concentration to one or both of the condensed phases is entirely downhill on the free-energy landscape.
The phase behavior that is observed on the timescales considered in the experiments is thus likely to be strongly influenced by the first condensed phase to form, especially considering the exceedingly slow kinetics at both high and low salt concentrations.
A potential consistent explanation is thus that the liquid-like droplets at \SI{1}{M} NaCl (Fig.~\ref{fig:PD}B) are the result of rapid phase separation to a condensed phase that incorporates both pNS and PLL and is stabilized by both homotypic and heterotypic interactions.
Moreover, it is possible that further phase separation into two immiscible condensed phases is not observed at \SI{1}{M} NaCl due to an insufficient driving force once the initial condensed phase has reached a local equilibrium.
By contrast, at low salt concentrations ($\rhoS \lesssim \rhoU$) and excess polyanion, the model predicts that the minimum-free-energy paths to both condensed phases traverse free-energy barriers, indicating that condensation must occur via nucleation.
This mechanism is more likely to produce distinct droplets of two immiscible condensed phases, leading to the direct observation of three-phase coexistence in Fig.~\ref{fig:PD}C.
Taken together, these various lines of evidence suggest that the phase behavior in Fig.~\ref{fig:PD}A--D is the product of both thermodynamic and kinetic effects, which in turn both depend on the balance between homotypic and heterotypic interactions at different macromolecular compositions and salt concentrations.

\subsection{Sequence-dependent NS partitioning into immiscible coacervates}
Finally, we asked whether the interplay among homotypic and heterotypic interactions can be used to assemble immiscible pNS+PLL coacervates enriched in distinct NS species.
Prior works have demonstrated the formation of immiscible pNS phases through the use of orthogonal palindromic sticky-end sequences that have no affinity for one another \cite{Sato2020, Jeon2020, tanase2025internal}.
However, in those cases, homotypic base pairing provided the sole driving force for phase separation.
Here, we instead consider the formation of multicomponent coacervates and investigate whether the heterotypic interactions involving the shared PLL component overwhelm the homotypic interactions, driving the distinct pNS phases to mix. 

\begin{figure}
    \centering
    \includegraphics[width=\columnwidth]{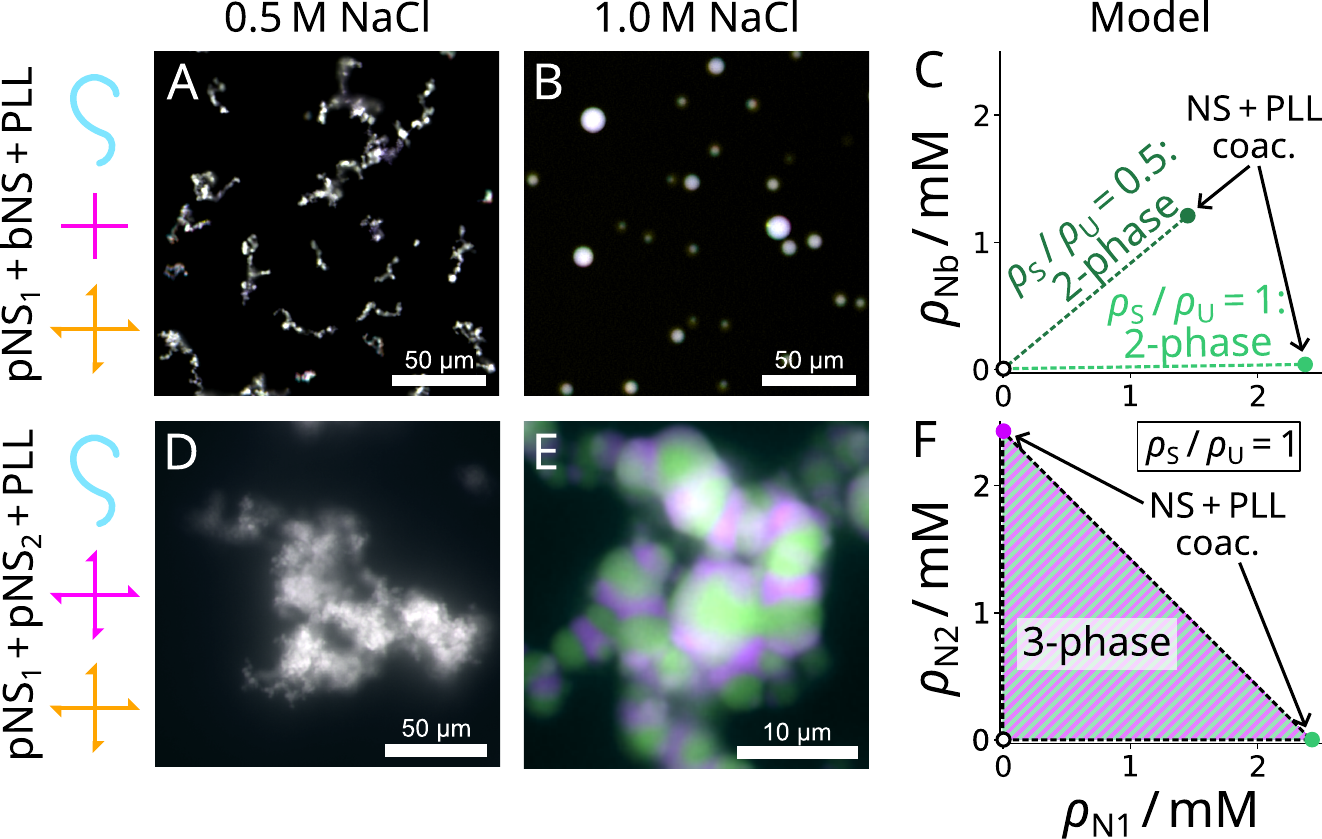}
    \caption{Variable-composition and orthogonal NS+PLL coacervates at \SI{25}{\celsius}.
      Mixtures of blunt NS (\SI{5}{\micro M} bNS, magenta), pNS (\SI{5}{\micro M}, yellow), and PLL (\SI{19.2}{\micro M}, cyan) form (A)~three-component, kinetically arrested coacervate gels at \SI{0.5}{M} NaCl (note that regions where magenta, cyan, and yellow colocalize appear gr
      ay; see also Fig.~\ref{fig:SI-multiphase-indiv}), and (B)~three-component liquid-like coacervates at \SI{1}{M} NaCl that are relatively enriched in pNS (Fig.~\ref{fig:SI-multiphase-partitioning}). (C)~Model predictions for tie lines and binodal points of dilute-dense equilibria in the plane of bNS concentration, $\rhoNb$, and pNS concentration, $\rhoNo$, for low  (dark green) and high  (light green) salt concentrations. At low salt, the dense phase has roughly equal NS partitioning ($\rhoNo\approx\rhoNb$), while at high salt concentrations, pNS partitions to a greater extent ($\rhoNo>\rhoNb$), matching the experiments.    
      (D)~Mixtures comprising two palindromic NSs with orthogonal sticky-end sequences, pNS$_1$ and pNS$_2$, also form three-component, kinetically arrested coacervates gels with PLL at \SI{0.5}{M} NaCl (\SI{5}{\micro M} pNS$_1$, magenta; \SI{5}{\micro M} pNS$_2$, yellow; \SI{19.2}{\micro M} PLL, cyan).
      (E)~At \SI{1}{M} NaCl, each pNS species forms liquid-like coacervates with PLL, but these condensed phases are immiscible with one another.
      (F)~At higher salt concentrations ($\rhoS = \rhoU$), the equilibrium theoretical phase diagram agrees with panel E in predicting three-phase coexistence, with the dilute phase (open point) coexisting with two distinct pNS+PLL coacervate phases enriched, respectively, in pNS$_1$ (magenta point) and pNS$_2$ (green point).
      Panels C and F show slices of the phase diagrams with the parent concentration fixed to the experimental concentrations in panels A,B and D,E, respectively.}
    \label{fig:multi}
\end{figure}

To this end, we first studied a mixture of PLL, pNS, and blunt NS (bNS) at a 1:\nicefrac{1}{2}:\nicefrac{1}{2} macromolecular charge ratio.
The bNS species has no sticky ends and thus does not contribute homotypic interactions.
At \SI{0.5}{M} NaCl, this mixture forms kinetically arrested coacervate gels incorporating all three macromolecular components, including both NS species, likely due to the dominant role of electrostatic interactions at this salt concentration (Fig.~\ref{fig:multi}A and Fig.~\ref{fig:SI-multiphase-indiv}).
At \SI{1}{M} NaCl, the three macromolecular components assemble into liquid-like droplets (Fig.~\ref{fig:multi}B and Fig.~\ref{fig:SI-multiphase-indiv}).
However, closer examination of the compositions of these droplets revealed greater partitioning of the pNS species at this salt concentration (Fig.~\ref{fig:SI-multiphase-partitioning}).
This finding is consistent with our previous conclusion that base pairing interactions are needed to stabilize droplets under these conditions (cf.~Fig.~\ref{fig:intro}D), since only the pNS species provides this homotypic contribution to the driving force for phase separation and can therefore be expected to be enriched in the multicomponent droplets to a greater extent.
This expectation is corroborated by our model, which predicts a shift in the equilibrium droplet NS composition from roughly equimolar pNS and bNS when $\rhoS < \rhoU$ to predominantly pNS when $\rhoS \gtrsim \rhoU$ (Fig.~\ref{fig:multi}C).
We emphasize that this compositional shift is a consequence of the electrostatic heterotypic interactions, since bNSs are predicted to be excluded from the condensed phase at all salt concentrations in the absence of PLL (Fig.~\ref{fig:SI-multiphase-theory}).

We then considered a mixture of PLL and two distinct pNS species with orthogonal palindromic sequences, pNS$_1$ (5'-{\fontfamily{qcr}\selectfont {\textit{CGATCG}}}-3') and pNS$_2$ (5'-{\fontfamily{qcr}\selectfont {\textit{GCTAGC}}}-3'), at a 1:\nicefrac{1}{2}:\nicefrac{1}{2} macromolecular charge ratio.
Although a single kinetically arrested coacervate gel was observed at \SI{0.5}{M} NaCl (Fig.~\ref{fig:multi}D and Fig.~\ref{fig:SI-multiphase-indiv}), phase separation led to the formation of two immiscible liquid-like coacervates at \SI{1}{M} NaCl (Fig.~\ref{fig:multi}E and Fig.~\ref{fig:SI-multiphase-indiv}).
PLL is enriched in both of these coacervates, but each is enriched in a single pNS species, with the orthogonal pNS species excluded.
The two condensed phases were found to adhere strongly to one another, presumably due to the shared PLL component, and to coarsen over time while remaining immiscible (Fig.~\ref{fig:SI-multiphase-coarsening}).
These observations are also consistent with our theoretical model, which predicts equilibrium three-phase coexistence between a dilute phase, a pNS$_1$+PLL condensed phase, and a pNS$_2$+PLL condensed phase when $\rhoS \simeq \rhoU$ (Fig.~\ref{fig:multi}F).
We therefore conclude that heterotypic interactions can indeed dominate over orthogonal homotypic interactions, but in a manner that is highly tunable with salt: at intermediate salt concentrations, the weakened heterotypic attractions allow for base pairing-driven demixing of the two pNS species while still supplying the driving force for pNS$_1$+PLL and pNS$_2$+PLL coacervation (Fig.~\ref{fig:SI-multiphase-theory}).
Importantly, the transition from a three-component coacervate to two two-component coacervates occurs due to the same cooperativity between electrostatics and base pairing that is necessary for pNS+PLL coacervation at intermediate salt concentrations.

\section{Conclusions}
Through a combination of experiments and theoretical calculations, we have shown how the interplay between heterotypic electrostatics and homotypic base pairing interactions govern the phase behavior of mixtures of nucleic acids and charged polypeptides.
By controlling the sticky-end sequences, salt concentration, temperature, and macromolecular composition, we have been able to distinguish contributions to the overall driving force for phase separation arising from these two interaction modalities.
These systems exhibit surprisingly complex phase behavior in the intermediate salt regime, where the magnitudes of the heterotypic and homotypic interactions are comparable due to the opposite effects of added salt on electrostatics and base pairing.

Our work reveals how electrostatic and base pairing interactions can either cooperate or compete with one another to shape the phase behavior of mixtures of nucleic acid and charged polypeptides.
On the one hand, these interaction modalities cooperate to increase the range of salt concentrations and temperatures at which coacervation can occur, most strikingly at high temperatures and high salt concentrations (Fig.~\ref{fig:intro}F,G and Fig.~\ref{fig:temp}C).
On the other hand, the observation of three-phase coexistence in mixtures with a single pNS species and PLL (Fig.~\ref{fig:PD}C) indicates that these interaction modalities can also compete, leading to the formation of distinct condensed phases (i.e., a pNS-rich phase and a spatially separate pNS+PLL phase) that are each stabilized primarily by one or the other type of interaction.
Competition between homotypic and heterotypic interactions is also responsible for the demixing of coacervate phases in mixtures of multiple pNS species with orthogonal sticky ends and PLL (Fig.~\ref{fig:multi}).
These conclusions are broadly compatible with the recent findings of Ohno \emph{et al.} \cite{ohno2025oligolysine}.

Our synthesis of experiments and theoretical calculations allows us to explain how the qualitative changes in both the equilibrium phase behavior and the nonequilibrium phase-separation kinetics arise from relatively subtle changes in the solution conditions.
By accounting for base pairing, electrostatic interactions, and the various entropic contributions of NSs, PLL, and microions, our model is able to reproduce all relevant experimental observations, including: the mixed homotypic and heterotypic interactions that stabilize pNS+PLL mixtures at intermediate salt concentrations (Fig.~\ref{fig:intro}I); the distinct temperature sensitivities of various mixtures (Fig.~\ref{fig:temp}D); the relative placement of the distinct regimes of pNS+PLL phase behavior in the salt/stoichiometry plane, and their dependence on a combination of equilibrium and kinetic factors (Fig.~\ref{fig:PD}); the existence of, and the conditions defining, three-phase equilibria in pNS+PLL mixtures (Fig.~\ref{fig:intro}I and Fig.~\ref{fig:PD}C,E); and the salt-dependent phase behavior of mixtures with three macromolecular components (pNS+bNS+PLL or pNS$_1$+pNS$_2$+PLL; Fig.~\ref{fig:multi}C,F).
Importantly, these equilibrium and nonequilibrium effects can be rationalized by our theoretical model without the need for fitting parameters, enabling mechanistic interpretations of the thermodynamic driving forces across a wide range of solution conditions.

These varied phase behaviors were all observed at easily accessible salt concentrations at room temperature.
In particular, for the specific pNS and PLL constructs studied in this work, the intermediate salt range at which the heterotypic and homotypic interactions become comparable to one another is located below \SI{1}{M} NaCl.
Our work therefore suggests that physiological solutions of nucleic acids and polypeptides are naturally poised to exhibit complex and easily tunable phase behaviors due to the comparable strengths of heterotypic and homotypic interactions arising from electrostatics and base pairing.

\section{Methods}
\subsection{NS formation}
\label{method:NS}
DNA oligomers were ordered from IDT with standard desalting purification (sequences given in Table \ref{tab:SI-sequences}). They were resuspended in \SI{10}{mM} Tris-HCl before being mixed at equimolar concentrations. The mixture was then annealed by heating at \SI{95}{\celsius} for 10 minutes and then cooled to \SI{4}{\celsius} at \SI{-0.5}{\celsius/min}. NSs were purified using \SI{0.22}{\micro m} PVDF and \SI{30}{kDa} regenerated cellulose centrifugal filters (Millipore Sigma and Amicon, respectively). Proper NS formation was confirmed using agarose gel electrophoresis (3\% agarose gel with TBE running buffer). NSs were stored in \SI{10}{mM} Tris-HCl at \SI{-20}{\celsius} between experiments.

\subsection{Sample preparation and visualization}
\SI{10}{mM} Tris-HCl, \SI{5}{M} NaCl, and water were combined and vortexed. Then monodisperse PLL (Alamanda Polymers, $n = 100$ unless otherwise stated) was added and pipette mixed. NSs, including 5\% that were fluorescently tagged, were incubated at \SI{50}{\celsius} dissolve any condensed phase. Finally, the NSs, lacking condensates, were added to PLL/salt solution with a cut pipette tip and the solution was pipette mixed three times, then immediately added to the flow cell for visualization. If multiple species of NSs were included, they were combined before being added to the salt-PLL mixture such that all NSs were added in a single step.

NS+PLL samples were measured in a flow channel constructed from parafilm sandwiched between polyacrylamide coated glass slides, prepared following Ref.~\cite{DeCamp2015}. Briefly, microscope slides and coverslips were  cleaned by sonicating with Hellmanex, ethanol, NaOH, and water. Slides were then incubated for 30 minutes in 3-(Trimethoxysilyl)propyl methacrylate, ethanol, and acetic acid. Slides were coated by leaving them in solution with acrylamide, TEMED, and ammonium persulfate overnight. Immediately before use, slides were rinsed with water and dried with pure N$_2$ gas. Parafilm was melted onto the slide with a soldering iron such that a narrow channel was formed. Samples were added to the channel with a cut pipette tip to reduce shearing. Channels were sealed using Norland Optical Adhesive  cured with a UV light. Samples were visualized using epifluorescent microscopy (Nikon Eclipse Ti2-E) and, unless otherwise stated, at \SI{25}{\celsius} (Oko UNO).

\subsection{Theoretical model}
\label{method:model}

We developed a theoretical model to predict equilibrium and kinetic phase behaviors of NS+PLL mixtures.
  Our approach accounts for the free energies of screened electrostatics and base pairing in aqueous solution, as well as the effects of translational entropy and excluded volume.
Each NS is represented as a single coarse-grained unit, referred to as a ``blob''.
Each PLL molecule is represented as a freely jointed chain of $N_\text{p}=10$ Kuhn-length blobs~\cite{shi2013control}.
Each blob carries a Gaussian charge distribution,
\begin{equation}
  q_i(r) = Z_i e \left(\frac{1}{2\pi\sigma_i^2}\right)^\frac{3}{2}\exp{(-r^2/2\sigma_i^2}),
\end{equation}
where we use the subscript $i$ to indicate the nanostar (N) and PLL (P) species.
A blob of species $i$ carries a total charge of $Z_i e$, where we use the bare charge values of one negative charge per NS base, except for the terminal 3' base that lacked a phosphate (and so $Z_N = -192)$, and one positive charge per PLL residue (and so $Z_P=10$).
The blob radius $\sigma_i$ describes the spatial range of the charge distribution.
Microcations (+) and microanions ($-$), including counterions and added monovalent salt, are treated as point particles.
We do not distinguish between these different sources of microions in the model.
The implicit solvent is treated as a continuum.
The number densities of these four ion species, $\rho_{\text{N}}$, $\rho_{\text{P}}$, $\rho_{+}$, and $\rho_{-}$, are constrained by the electroneutrality condition
\begin{equation}
  Z_{\rm N} \rho_{\rm N}+\NP Z_{\rm P} \rho_{\rm P}+\rho_+-\rho_- = 0.
\end{equation}

We evaluate the system free energy within a mean-field approximation by adapting the DLVO theory~\cite{derjaguin1941theory,verwey1947theory} for colloidal particles to soft blobs.
We then account for associative interactions between sticky ends using Wertheim's first-order perturbation theory~\cite{wertheim1984fluids}.
Because the system free energy is derived from the classical partition function of the blob mixture using classical density functional theory, this approach treats the screened electrostatic and associative interactions in a consistent manner.
This consistency between the treatment of electrostatic and base pairing interactions is crucial for the prediction of overlapping upper and lower critical salt concentrations for npNS+PLL coacervation and pNS associative phase separation, respectively.
Complete details are provided in Sections S1--S5.

The total free-energy density of the mixture can be written as
\begin{equation}
  f = f_\text{poly}+f_\text{micro}+f_\text{self}+f_\text{assoc}.
\end{equation}
The first term represents the polyion reference state, accounting for the ideal translational entropy of a mixture of NSs and PLL, as well as excluded volume interactions.
Following our prior work~\cite{li2023interplay}, this term is
\begin{align}
  \beta f_{\text{poly}} &= \rho_{\text{N}} \ln (\rhoN\vN) + \rhoP \ln \left(\frac{\rhoP\NP\vP}{1 - \rhoN\vN} \right) \nonumber \\
  &\quad + \left(\frac{1 - \rhoN\vN}{\vN}\right) \ln (1 - \rhoN\vN) \nonumber \\
  &\quad + \left(\frac{\phi_0}{\vP}\right) \ln \left( \frac{\phi_0}{1 - \rhoN\vN} \right),
\end{align}
where $\beta\equiv(k_\text{B}T)^{-1}$, $v_\text{N} = 4\pi\sigma_{\text{N}}^3/3 = 524 \text{ nm}^3$ and $v_\text{P} = 4\pi\sigma_{\text{P}}^3/3 = 0.904\text{ nm}^3$ are the volumes of N and P blobs, respectively, and $\phi_0 \equiv 1 - \rhoN\vN - \rhoP\NP\vP$ is the solvent-occupied volume fraction.
In the absence of electrostatic interactions, microions also make ideal (i.e., translational entropy) and excluded-volume contributions,
\begin{align}
  \beta f_\text{micro} &= \rho_+ \ln (\rho_+ v_0)+\rho_- \ln (\rho_- v_0) \nonumber \\
  &\quad - (\rho_++\rho_-)\ln{(1 - \eta)}.
\end{align}
The volume $v_0$ of the microion point particles is included for dimensional consistency, although its value does not affect phase-coexistence calculations.
The last term in $f_\text{micro}$ accounts for the exclusion of microions from sterically inaccessible regions within the polyion blobs~\cite{widom1963some}.
The total volume fraction of the system that is inaccessible to microions is 
\begin{equation}
  \eta = f_\text{N}\rhoN\vN+f_\text{P}\rhoP\NP\vP,
\end{equation}
where $f_{\text{N}}\approx0.01$ and $f_{\text{P}}\approx0.5$ are the inaccessible fractions of the N and P blob volumes, respectively (Table~\ref{tab:parameters}).

At the mean-field level, the electrostatic interactions among all the solvated species reduce to a ``self-energy'' resulting from the formation of the electrical double layer, in which each polyion is immersed in an oppositely charged cloud of microions. The solvent is represented as a continuum with a temperature-dependent dielectric constant, $\epsilon(T)$, that accounts for its thermal reorganization\cite{chen2022driving,archer1990dielectric}. This is specifically embodied in the Bjerrum length, $l_B \equiv e^2 / \epsilon k_{\text{B}}T \approx \SI{0.71}{nm}$, which is nearly independent of temperature due to the fact that $\epsilon(T) \times T$ is approximately constant in aqueous solution. For simplicity, we treat $l_{\text{B}}$ as a constant in the model.
The electrostatic contribution to the free-energy density is thus
\begin{align}
  \beta f_\text{self} &= -\frac{\kappa l_\text{B}}{2}\left[\rhoN Z^2_\text{N}  \exp{(\kappa^2\sigma_\text{N}^2)}\,\text{erfc}(\kappa \sigma_\text{N})\right. \nonumber \\
    &\quad \left.+\NP\rhoP Z^2_\text{P} \exp{(\kappa^2\sigma_\text{P}^2)}\,\text{erfc}(\kappa \sigma_\text{P})\right],
\end{align}
where
$\kappa = \left[ 4\pi l_\text{B} \left( \rho_+ + \rho_- \right) \right]^{1/2}$ is the salt-dependent inverse Debye screening length.

The associative contribution to the free energy is~\cite{wertheim1984fluids}
\begin{equation}
 \beta f_{\text{assoc}} = m \rhoN \left(\ln X - \frac{X}{2} + \frac{1}{2}\right),
\end{equation}
where $m=4$ is the number of sticky ends per pNS and $X$ is the probability that an individual sticky end on a pNS is unbound,
\begin{equation}
  X = \frac{-1 + \sqrt{1 + 4 \rhoN \vN \Delta}}{2 \rhoN \vN \Delta}.
\end{equation}
The dimensionless sticky-end association strength, $\Delta$, is related to the hybridization free energy, $\Delta G$, of sticky-end oligonucleotides that are free in solution via
\begin{equation}
    \Delta = \frac{m\Omega}{4\pi} \exp({-\beta \Delta G})\, v_\text{N}^{-1} \cdot \text{M}^{-1},
\end{equation}
where $\Omega = {\pi}/{4}$ is the solid angle accessible to a sticky-end on a pNS blob~\cite{hegde2024competition}, and the oligonucleotide hybridization free energies are reported relative to a standard concentration of \SI{1}{M}~\cite{santalucia2004thermodynamics}.
Standard $\Delta G^\circ$ values were computed as a function of temperature at \SI{1}{M} NaCl using NUPACK~\cite{NUPACK2020}.
These standard values were reduced by $3.06 k_\text{B}T$ to avoid double counting of salt effects, which are also accounted for implicitly in the NUPACK predictions.
An experimentally validated empirical salt correction~\cite{santalucia2004thermodynamics} was then applied to obtain a salt-dependent association strength,
\begin{equation}
  \Delta = \Delta_0 \left(\rho_+ \cdot \text{M}^{-1}\right)^{\alpha},
\end{equation}
where $\Delta_0 = \Delta(\Delta G^\circ)$ is the association strength at \SI{1}{M} monovalent cation concentration and $\alpha = 1.15$.
The association contribution to the free energy therefore depends explicitly on both temperature and salt concentration via the association strength $\Delta$. 

Phase diagrams were computed by first finding the convex hull of the free-energy density as a function of $\rhoN$, $\rhoP$, and $\rhoS$ and then refining the boundaries of the coexistence regions and the predicted tie lines by explicitly solving for equal chemical potentials for all species and equal pressures in all coexisting phases.
Kinetic pathways were then predicted between an initial homogeneous solution at the parent concentration and each of the (meta)stable phases by calculating the minimum-free-energy path on the free-energy landscape using the string method~\cite{string2007}.
Complete details of these numerical methods are provided in Sections S1--S5.

Overall, the model predicts phase diagrams and kinetic pathways using a set of parameters that are determined based on their physical values or prior measurements, leaving no fitting parameters for comparison with the experiments.
  We emphasize that these parameters account for both energetic and entropic contributions to base pairing (through $\Delta G$) and to solvent behavior (through $l_B$), along with the size and charge of the polyions.
Due to the mean-field approximations employed in the model, the salt dependence of the predicted phase diagrams is not quantitative; however, by making comparisons based on the upper and lower transition salts, $\rho_{\text{U}}$ and $\rho_{\text{L}}$, respectively, we find agreement between the equilibrium and kinetic phase behaviors predicted by the model and the distinct regimes observed in the experiments across a wide range of physical conditions.

\begin{acknowledgements}
  Research reported in this publication was supported by the the National Science Foundation under award number DMR-2113302 to OAS and DMR-2143670 to WMJ; by an award from the W.M. Keck Foundation to OAS; and by the National Institute of General Medical Sciences of the National Institutes of Health under award number R35GM155017 to WMJ.

  \vskip1ex

  See the Supporting Information for Table S1 and Figs.~S1--S13, which are referenced in the main text.
  The Supporting Information also contains Sections~S1--S5, which describe the derivation and parametrization of the theoretical model, along with the dominance analysis, and contains additional supplementary Table S2 and Figs.~S14--S16.

\end{acknowledgements}

\clearpage
\onecolumngrid

\setcounter{section}{0}
\setcounter{equation}{0}
\setcounter{figure}{0}
\setcounter{table}{0}
\renewcommand{\thefigure}{S\arabic{figure}}
\renewcommand{\thetable}{S\arabic{table}}
\renewcommand{\theequation}{S\arabic{equation}}
\renewcommand{\thesection}{S\arabic{section}}

\newcommand{\longequal}{{=\!\!=}}
\newcommand{\tmmathbf}[1]{\ensuremath{\boldsymbol{#1}}}
\newcommand{\tmop}[1]{\ensuremath{\operatorname{#1}}}

\newcommand{\figref}[1]{Figure~\ref{#1}}
\renewcommand{\eqref}[1]{Eq.~(\ref{#1})}
\newcommand{\secref}[1]{Sec.~\ref{#1}}

\newcommand{\kB}{k_{\text{B}}}

\section*{Supplementary Information}

\textbf{Supplementary Information for ``Cooperation and competition of base pairing and electrostatic interactions in mixtures of DNA nanostars and polylysine''}

\section*{Supplementary Table}
\renewcommand{\arraystretch}{1.3}

\begin{longtable}{|c|>{\ttfamily\smaller}l|}
\caption{DNA sequences used in experiments. Fluorescent tags were included on the first oligo of each NS. 
}
\\
\hline
\textbf{NS-oligo \#} & \textbf{Sequence} \\
\hline
\endhead

\hline
\endfoot

\hline
\endlastfoot

pNS-1 & CGA TCG ACG CTG CAA CTG GAG GAT ACG AAG CCG TGG CAA GTC AGG TGC G \\
\hline
pNS-2 & CGA TCG ACG GCT CAG TCG GTT TCC GAG AAC GTA TCC TCC AGT TGC AGC G \\
\hline
pNS-3 & CGA TCG ACG AGC GTT GGA CAT GTA TCG AAC TCG GAA ACC GAC TGA GCC G \\
\hline
pNS-4 & CGA TCG ACG CAC CTG ACT TGC CAC GGC AAC GAT ACA TGT CCA ACG CTC G \\
\hline
npNS-1 & GCG CGC GGG CAG TTC TCC GCT TCG CTC GCT CGT CGG CCG CGC \\
\hline
npNS-2 & ACC TTC CTG CGC GCG GCT CGC ATT CCC GCG TTG GCA ATG GGC CTG GCG GCG CTT TCC AGA \\
\hline
npNS-3 & GCG CCG CCA GGC CCA TTG CCT TGC GGA GAA CTG CCC GCG CGC TTT CCA GA \\
\hline
npNS-4 & ACC TTC CTG CGC GCG GCC GAC GAG CGA GCG TTC GCG GGA ATG CGA GCC GCG C \\
\hline
Blunt NS-1 & TCG CTG CAA CTG GAG GAT ACG AAG CCG TGG CAA GTC AGG TGC G \\
\hline
Blunt NS-2 & TCG GCT CAG TCG GTT TCC GAG AAC GTA TCC TCC AGT TGC AGC G \\
\hline
Blunt NS-3 & TCG AGC GTT GGA CAT GTA TCG AAC TCG GAA ACC GAC TGA GCC G \\
\hline
Blunt NS-4 & TCG CAC CTG ACT TGC CAC GGC AAC GAT ACA TGT CCA ACG CTC G \\
\hline
pNS2-1 & GCT AGC ACA CCG CCC GGG CAG AAC AGG AAC GGT GAT ATC CCG GGC CTC G \\
\hline
pNS2-2 & GCT AGC ACT TCG CCC GGG TGC TAA GAG AAC CTG TTC TGC CCG GGC GGT G \\
\hline
pNS2-3 & GCT AGC AGC CTT CCC GGG AGC GCT CGC AAC TCT TAG CAC CCG GGC GAA G \\
\hline
pNS2-4 & GCT AGC ACG AGG CCC GGG ATA TCA CCG AAG CGA GCG CTC CCG GGA AGG C \\
\label{tab:SI-sequences}
\end{longtable}

\newpage
\section*{Supplementary Figures}

\begin{figure}[h!]
 \centering
  \includegraphics[width=\textwidth]{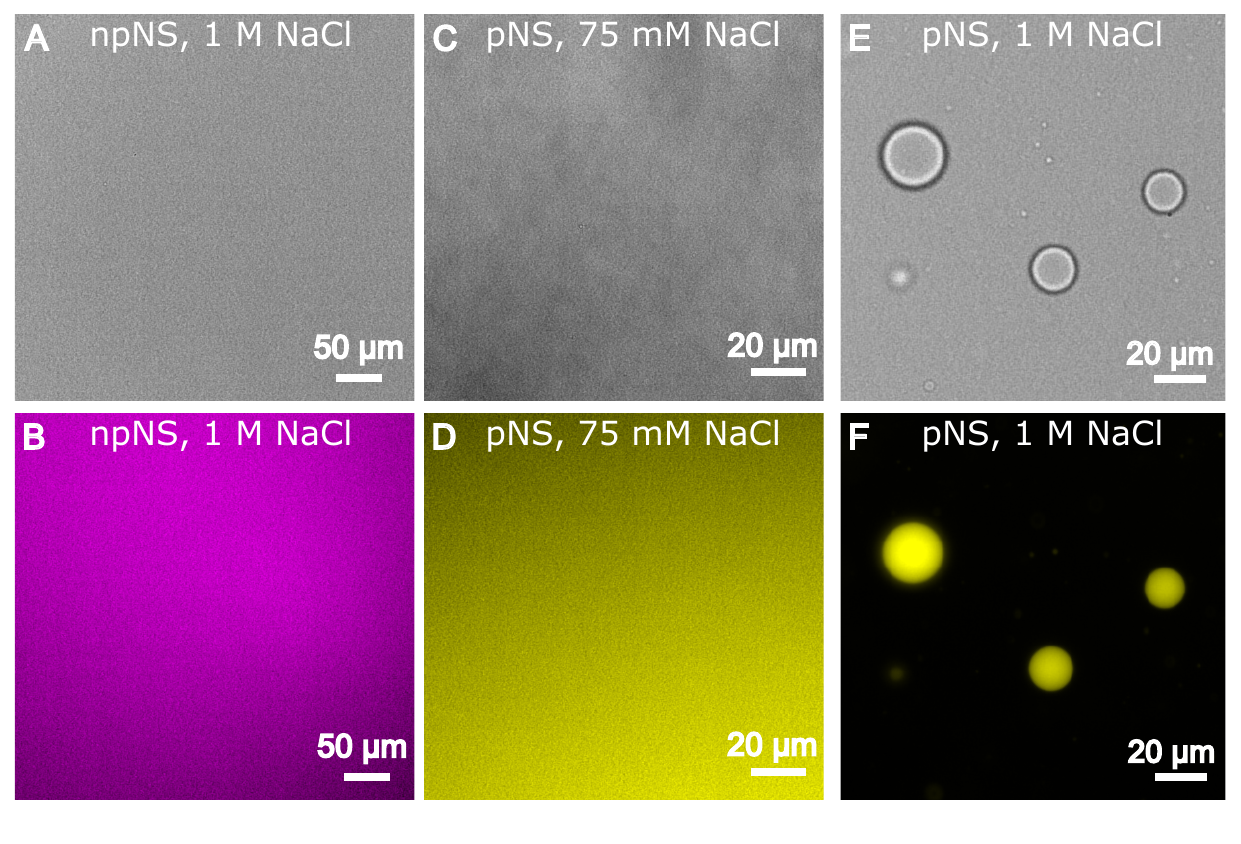}
  \caption{Representative images of NS phase behavior in the absence of PLL. (A,B)~npNSs (magenta) remain well mixed at \SI{1}{M} NaCl, shown in (A)~brightfield and (B)~fluorescence. (C,D)~pNSs (yellow) do not phase separate at \SI{75}{mM} NaCl, shown in (C)~brightfield and (D)~fluorescence. (E,F)~pNSs phase separate at \SI{1}{M} NaCl, shown in (E)~brightfield and (F)~fluorescence. All conditions contain \SI{10}{\micro M} NS.}
  \label{fig:SI-NSalone}
\end{figure}

\clearpage

\begin{figure}[p]
 \centering
  \includegraphics[width=0.6\linewidth]{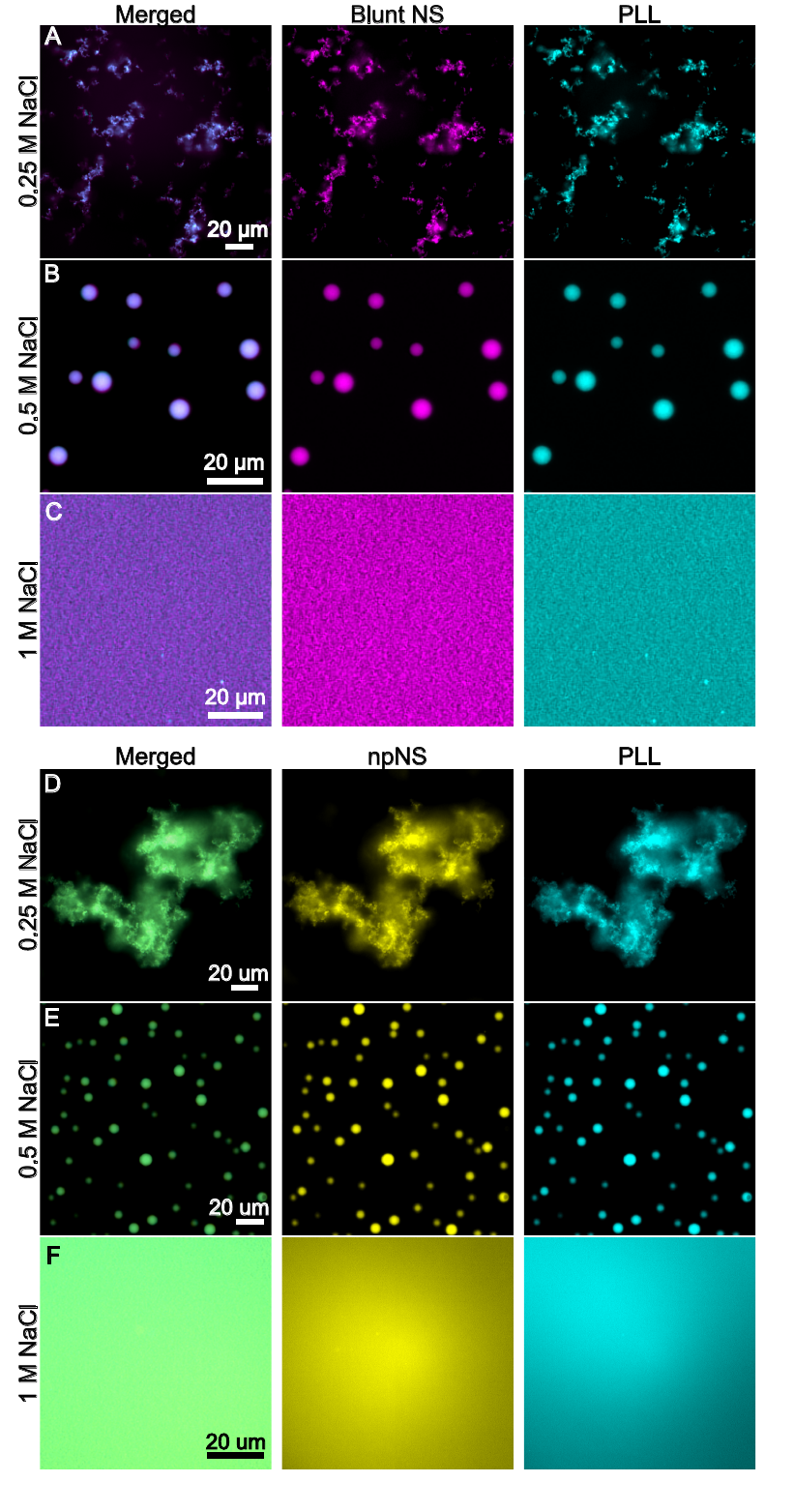}
  \caption{npNS+PLL and blunt NS+PLL phase separation. Blunt NS (magenta) and PLL (cyan) (A)~form kinetically trapped gels at \SI{0.25}{M} NaCl, (B)~form drops at \SI{0.5}{M} NaCl, and (C)~remain well-mixed at \SI{1}{M} NaCl. Similarly, npNSs with non-palindromic sticky ends (yellow) and PLL (D)~form kinetically trapped gels at \SI{0.25}{M} NaCl, (E)~form drops at \SI{0.5}{M} NaCl, and (F)~remain well-mixed at \SI{1}{M} NaCl. All mixtures contain 10 $\mu$M NS and $19.2~\mu$M PLL.}
  \label{fig:SI-npPhaseDiagram}
\end{figure}

\clearpage

\begin{figure}[p]
    \centering
    \includegraphics[width=0.7\linewidth]{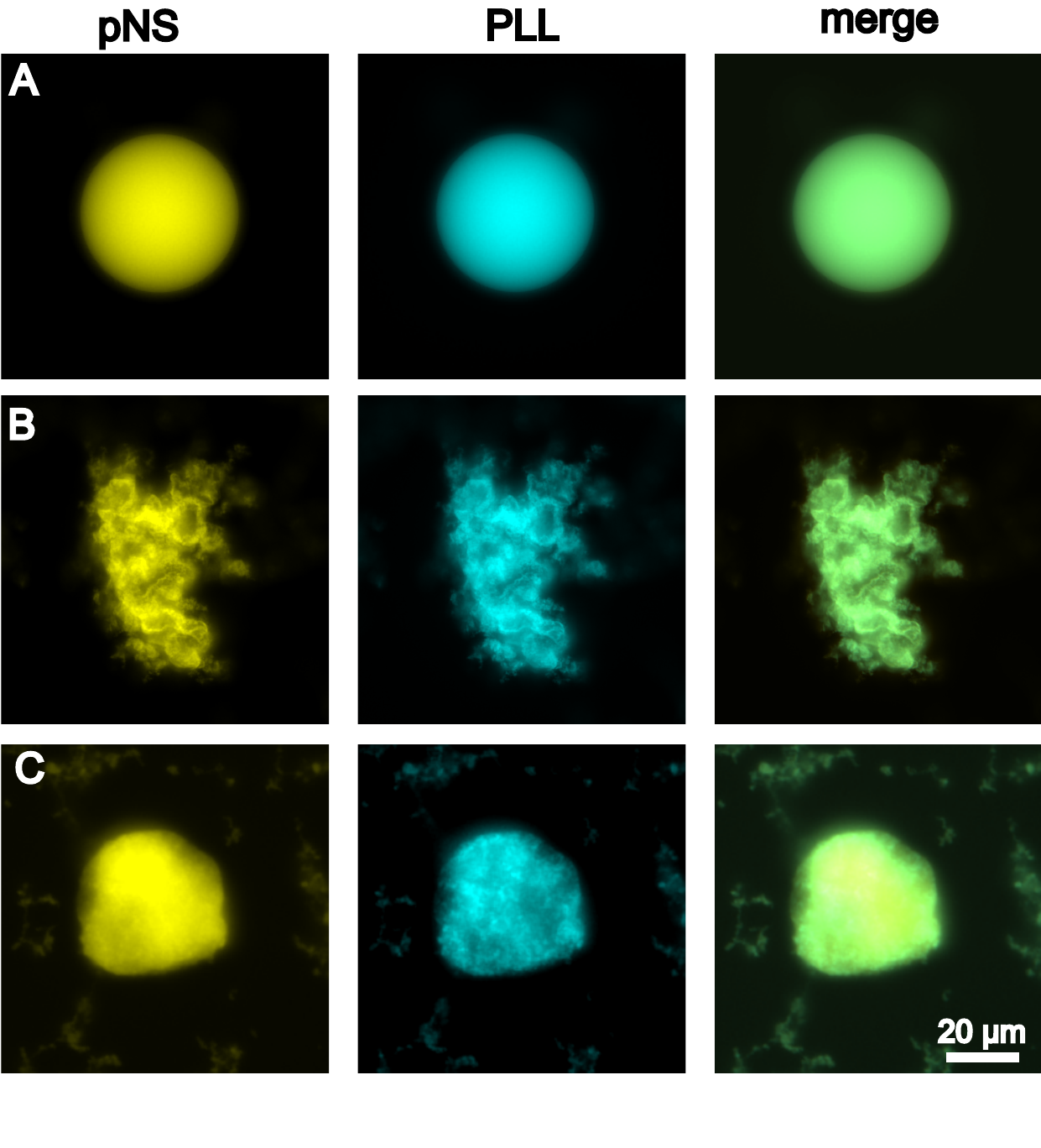}
    \caption{2-channel fluorescent images of various pNS+PLL mixtures after 48 hours at room temperature. The condensed phase in each situation is similar to the images taken shortly after mixing, and in the corresponding condition, as shown in Fig.~3 in the main text; this indicates a lack of aging, and near-equilibrium behavior, in these conditions. (A)~\SI{1}{M} NaCl at 1:1 macromolecular charge ratio, (B)~\SI{0.5}{M} NaCl at \nicefrac{1}{2}:1 macromolecular charge ratio, and (C)~\SI{0.25}{M} NaCl \nicefrac{1}{4}:1 macromolecular charge ratio.}
    \label{fig:SI-Stability}
\end{figure}

\clearpage

\begin{figure}[p]
\centering
\includegraphics[width=0.9\linewidth]{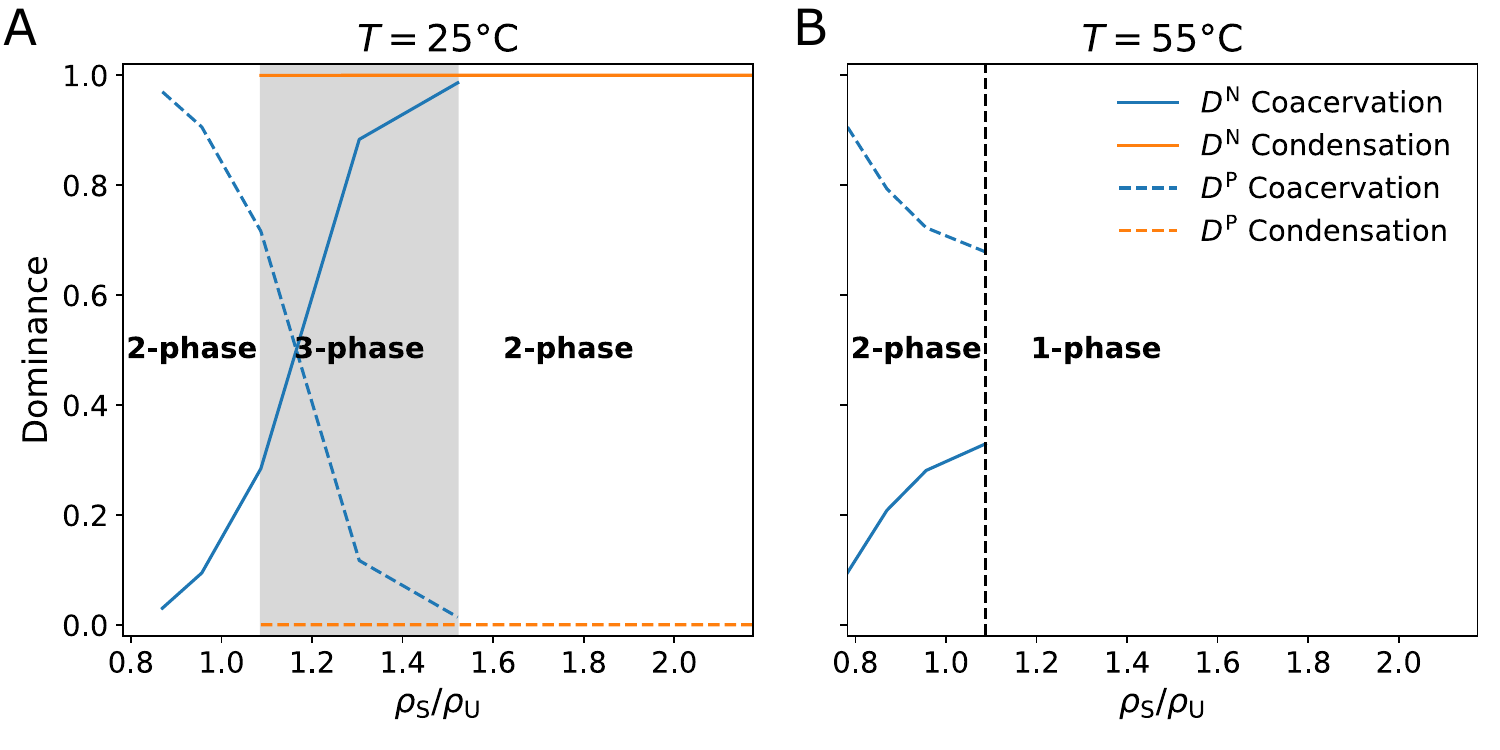}
\caption{Dominance analysis for macromolecularly charge-balanced pNS+PLL mixtures at (A) \SI{25}{\celsius} and (B) \SI{55}{\celsius}.  The parent concentrations are $\rhoN = \SI{10}{\micro M}$ and $\rhoP = \SI{19.2}{\micro M}$.  The dominance of pNS, $D^{\text{N}}$ (solid lines), and PLL, $D^{\text{P}}$ (dashed lines), are shown for both pNS-dominated condensates (orange lines) and pNS+PLL coacervates (blue lines).  Within the three-phase region, dominance is computed using the tie-lines connecting the dilute phase to either condensed phase.  The dominance metric for the microcation, $D^+$, is not shown since $|D^+| < 8\times10^{-3}$ in these conditions.  See \secref{sec:SI-dominance} for a description of the dominance analysis and a detailed discussion of these results.}
\label{fig:SI-dominance}
\end{figure}

\clearpage

\begin{figure}[p]
    \centering
    \includegraphics[width=0.7\linewidth]{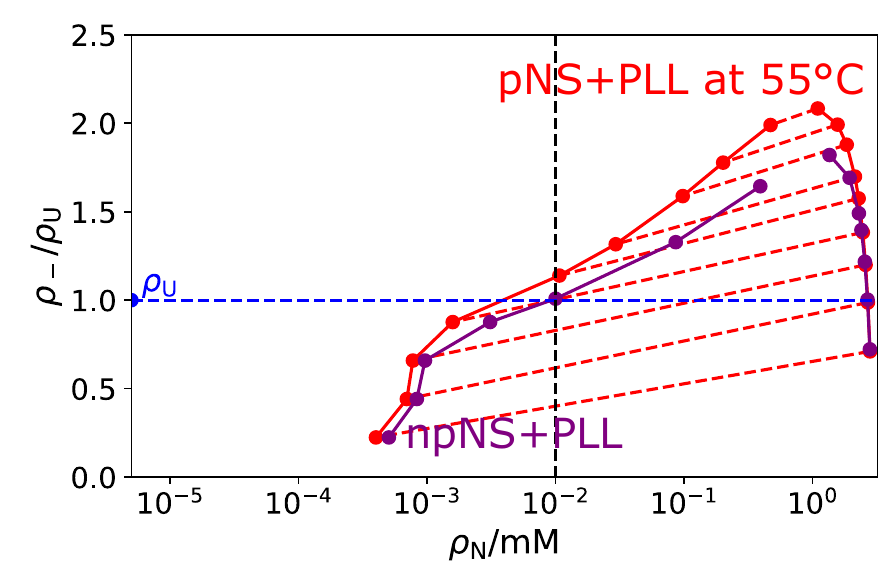}
    \caption{Comparison of predicted phase diagrams for the pNS+PLL and npNS+PLL systems at \SI{55}{\celsius}. The increased upper transition salt concentration indicates cooperativity between electrostatic and base pairing interactions.}
    \label{fig:SI-shift}
\end{figure}

\clearpage
\begin{figure}[p]
\centering
  \includegraphics[width=\textwidth]{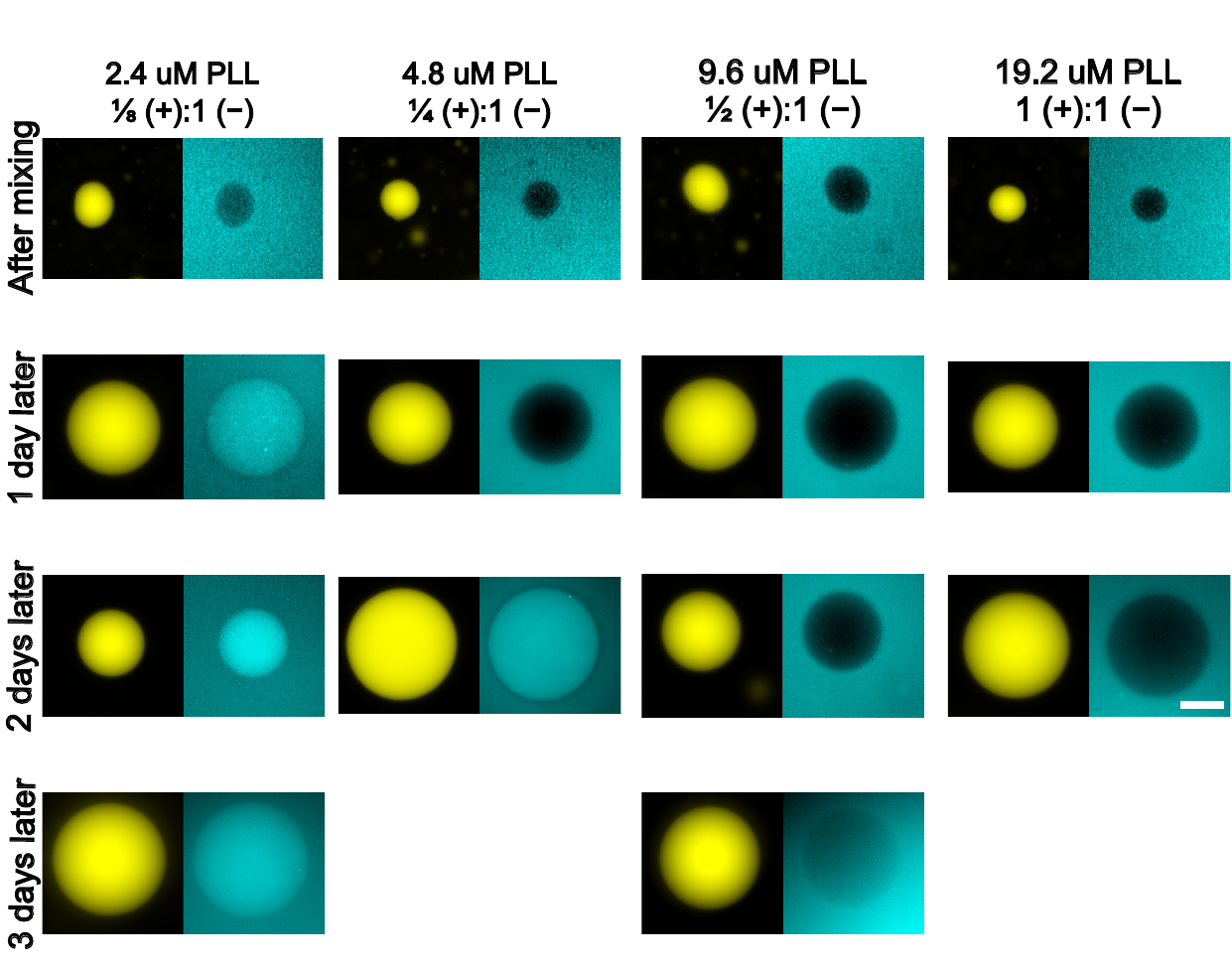}
  \caption{Fluorescent images of pNS (yellow) + PLL (cyan) mixtures at \SI{2.5}{M} NaCl over time, showing  slow dynamics of PLL partitioning, with timescales of hours to days.   
  This figure shows that the timescale is sensitive to PLL concentration, with enrichment occurring more quickly in systems that are further from macromolecular charge balance. See Fig.~\ref{fig:SI-slow-dynamicsPLL20} for data showing the sensitivity of enrichment timescale to PLL length.
It is surprising that such slow timescales are observed at very high salt, as one might have guessed that the reduced electrostatic interactions would permit PLL to move relatively quickly through a DNA matrix. 
We speculate that a coacervate membrane might be forming around the pNS droplets shortly after their formation, and that membrane might impede further PLL penetration.
This hypothesis could also potentially explain why longer PLL penetrates more slowly, as the membrane might be more stable due to more extensive entanglement of longer PLL with the DNA.
  Scale bar = \SI{20}{\micro m}.}
  \label{fig:SI-slowDynamics}
\end{figure}

\clearpage

\begin{figure}[p]
\centering
  \includegraphics[width=\textwidth]{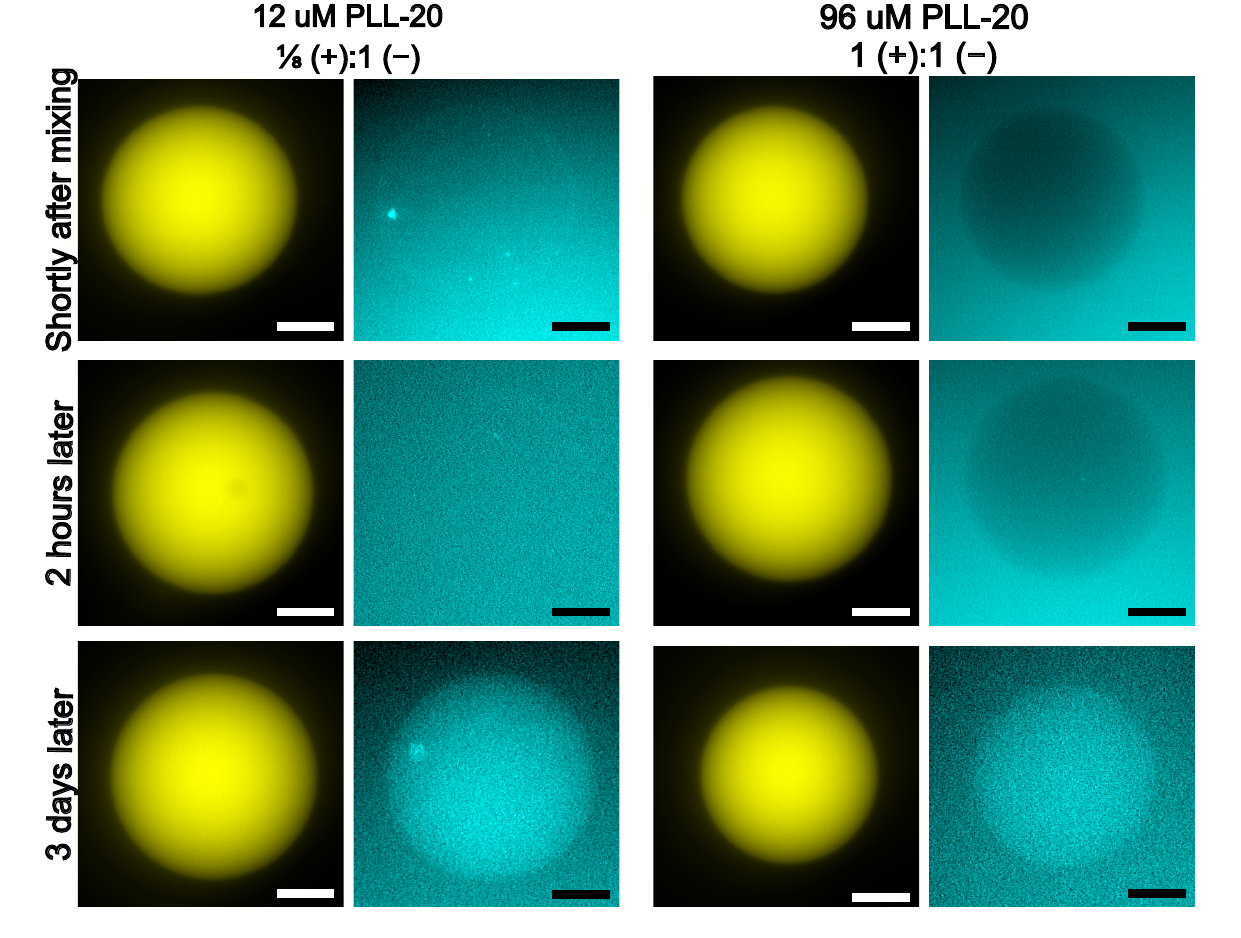}
  \caption{PLL$_{20}$ (20 residues) partitions faster than PLL$_{100}$ (100 residues) at \SI{2.5}{M} NaCl. \textit{Left:}t a \nicefrac{1}{8}:1 macromolecular charge ratio, PLL$_{20}$ is roughly equally concentrated in the dense and dilute phases after two hours, then becomes slightly enriched after three days. \textit{Right:} At a 1:1 macromolecular charge ratio, PLL$_{20}$ is excluded shortly after mixing and two hours later, and then becomes slightly enriched within three days. Scale bars = \SI{10}{\micro m}.}
  \label{fig:SI-slow-dynamicsPLL20}
\end{figure}

\clearpage

\begin{figure}[p]
\centering
 \includegraphics[width=0.75\textwidth]{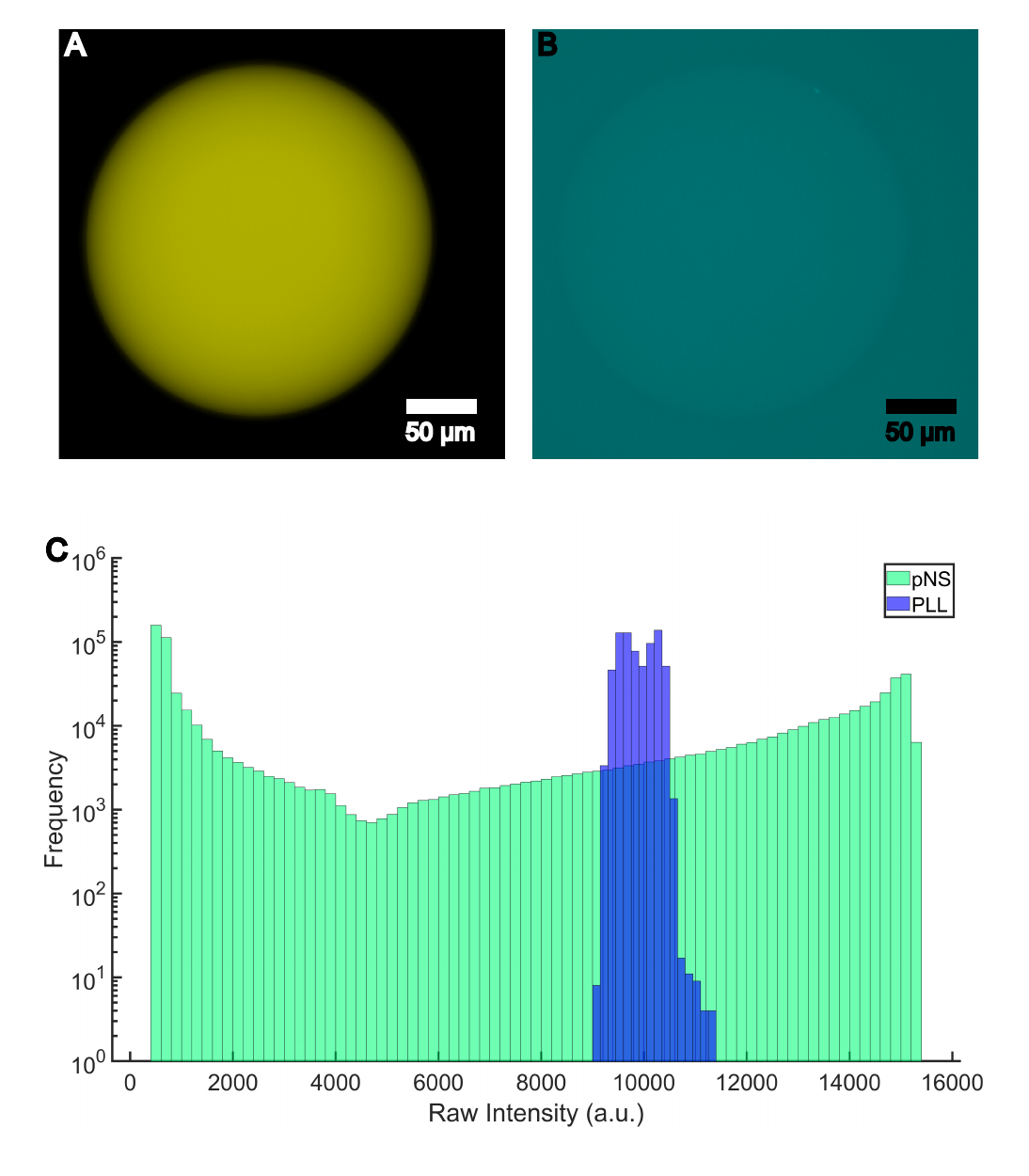}
  \caption{The partition coefficients of pNSs are higher than for PLL
  The pNSs are more strongly enriched in droplets than PLL after three days of incubation at \SI{2.5}{M} NaCl. Representative images of (A)~pNS (yellow) and (B)~PLL (cyan) at a 1:\nicefrac{1}{4} macromolecular charge ratio (i.e., excess pNS).
    (C)~Histogram of raw intensities of all pixels (i.e., both inside and outside of the condensed phase) for the epifluorescent images shown in (A) and (B). The broad distribution in the pNS channel arises from the dilute/dense intensity variation caused by the high partitioning into the droplet, along with the geometrical variation in intensity arising from epifluorescent imaging of a sphere. The narrow distribution in the PLL channel occurs because there is little variation in concentration inside and outside of the droplet.
  }
\label{fig:SI-2.5MPartitioning}
\end{figure}

\clearpage

\begin{figure}[p]
    \centering
    \includegraphics[width=0.75\linewidth]{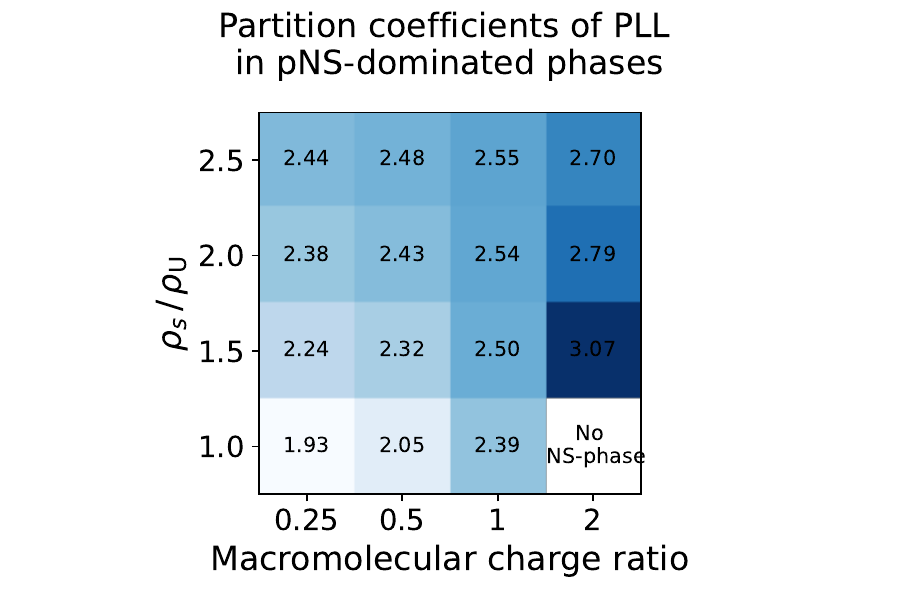}
    \caption{Predicted partition coefficients of PLL in the pNS+PLL systems. The partition coefficients of PLL suggest preferential partitioning into pNS-dominated condensates, even at high salt concentrations.}
    \label{fig:SI-pc}
\end{figure}

\clearpage

\begin{figure}[p]
\centering
\includegraphics[width=\textwidth]{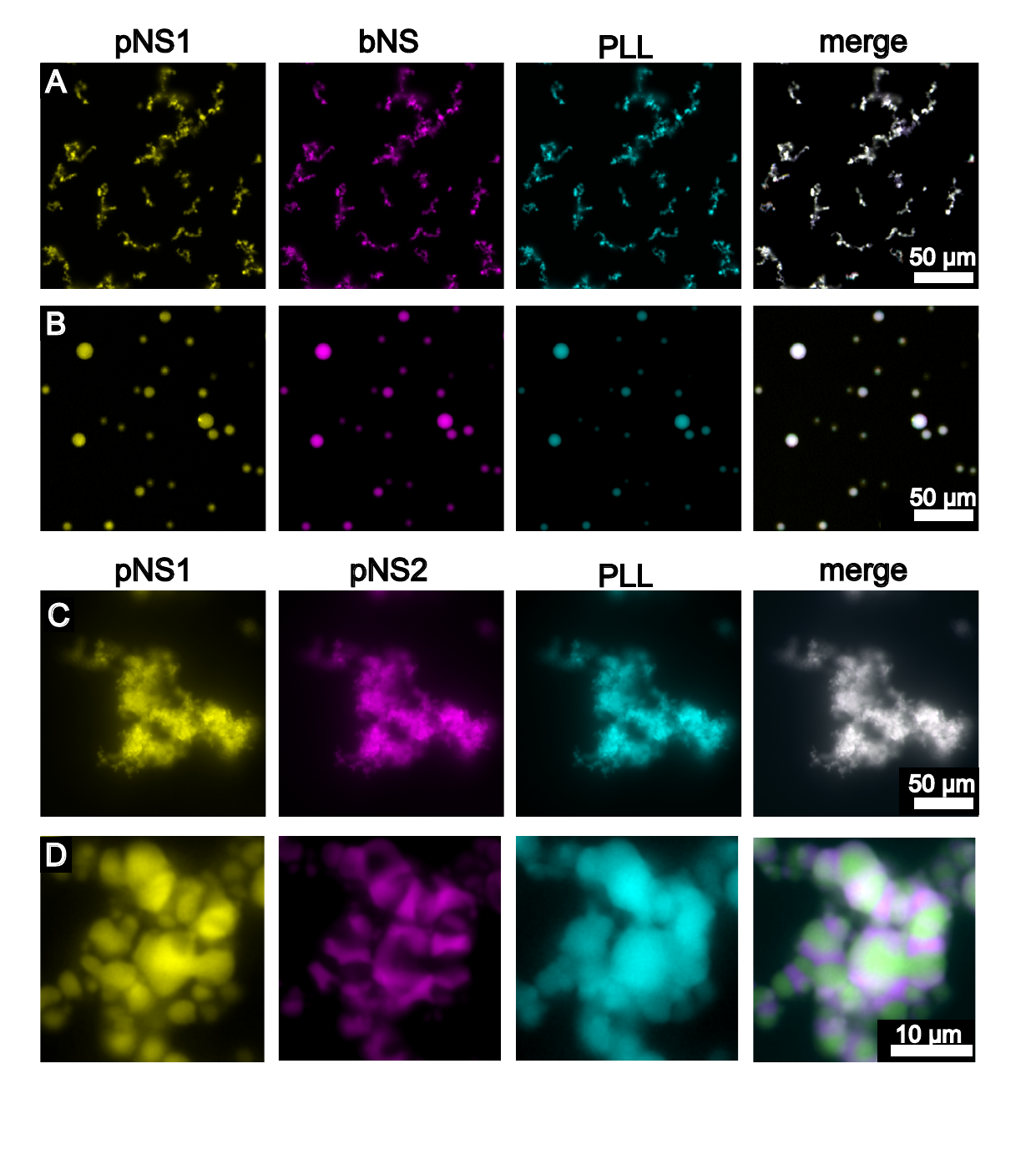}
\caption{Fluorescent imaging of multiphase structures. PLL (cyan), pNS (yellow) and bNS (magenta) form (A)~three-component aggregates at \SI{0.5}{M} NaCl and (B)~three-component liquid drops at \SI{1}{M} NaCl. (C)~Similarly, orthogonal pNS$_1$ (yellow) and pNS$_2$ (magenta), along with PLL (cyan), form three-component aggregates at \SI{0.5}{M} NaCl. (D)~However, at \SI{1}{M} NaCl, the pNS$_1$+PLL phase demixes from the pNS$_2$+PLL phase.}
\label{fig:SI-multiphase-indiv}
\end{figure}

\clearpage

\begin{figure}[p]
\includegraphics[width=\textwidth]{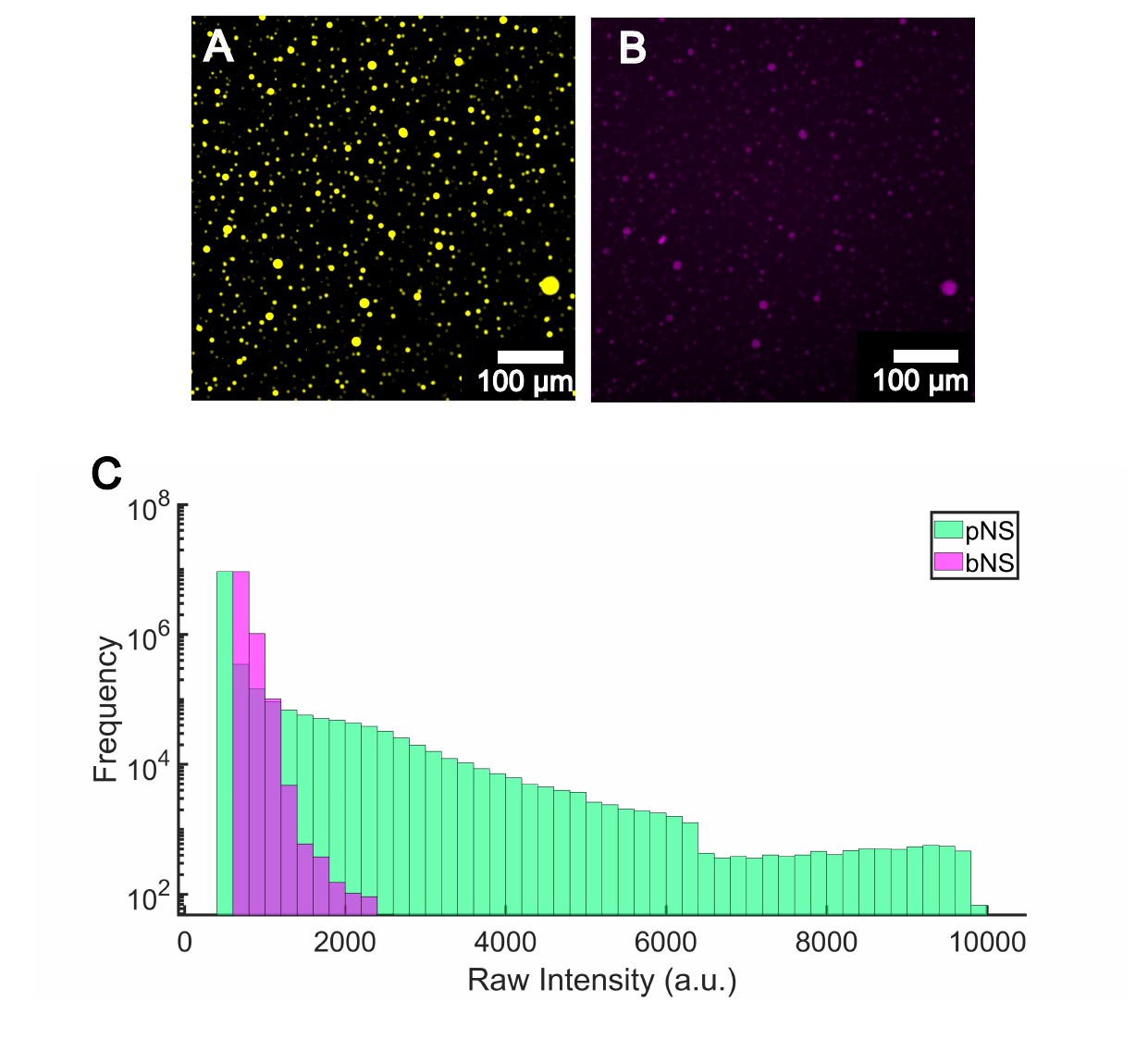}
  \caption{Fluorescent images of droplets in a 3-component solution containing \SI{5}{\micro M} pNS, \SI{5}{\micro M} bNS, and \SI{19.2}{\micro M} PLL, at \SI{1}{M} NaCl. The images show (A) the pNS channel (yellow), and (B) the bNS channel (magenta).  The pixel intensities of (A)~and (B)~use the same scale to show that pNSs are preferentially enriched relative to bNS in multicomponent droplets. (C)~Raw pixel intensities, from three indepenently prepared replicates, of pNS and bNS channels show a relatively constant bNS signal due to little difference in dense/dilute concentrations, and a wide variation in pNS signal due to the high partitioning along with the geometrical issues associated with imaging polydisperse droplets using an epi-fluorescent microscope.
  \label{fig:SI-multiphase-partitioning}
  }
\end{figure}

\clearpage

\begin{figure}[p]
\includegraphics[width=0.8\linewidth]{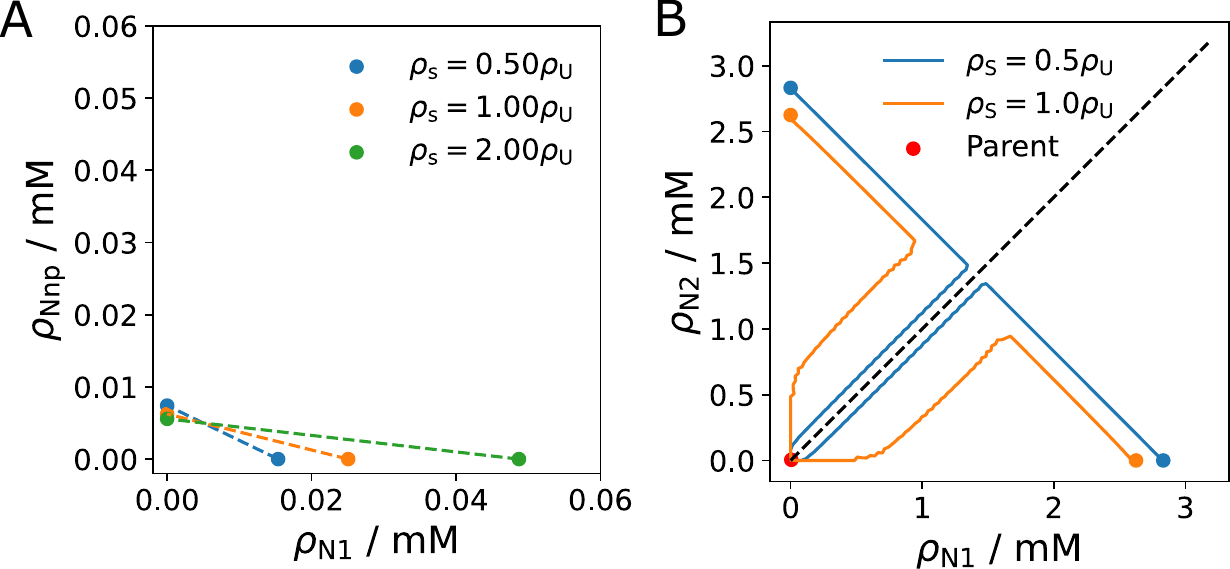}
  \caption{Theoretical predictions for multi-NS systems. (A)~ The predicted tie lines for systems containing equal amounts of pNS and npNS (parent concentration $\rho_{\text{N1}} = \rho_{\text{Nnp}} = 5\mu \text{M}$, $\rho_{\text{P}} = 0$) at three salt concentrations.  The blunt NS (bNS) is always excluded from the condensed phase regardless of the salt concentration, which differs qualitatively from the behavior shown in Fig.~4C in the main text.  (B)~Minimum-free-energy path predictions for pNS$_1$+pNS$_2$+PLL systems (parent concentration $\rho_{\text{N1}} = \rho_{\text{N2}} = 5\mu \text{M}$, $\rho_{\text{P}} = 19.2\mu \text{M}$) at two salt concentrations. At low salt concentration, the MFEP passes through an intermediate state enriched in both NS species, where the system may become kinetically trapped. At high salt concentration, distinct phases, each enriched in one or the other pNS species, form almost instantaneously.}
  \label{fig:SI-multiphase-theory}
\end{figure}

\clearpage

\begin{figure}[h!]
   \centering
  \includegraphics[width=\textwidth]{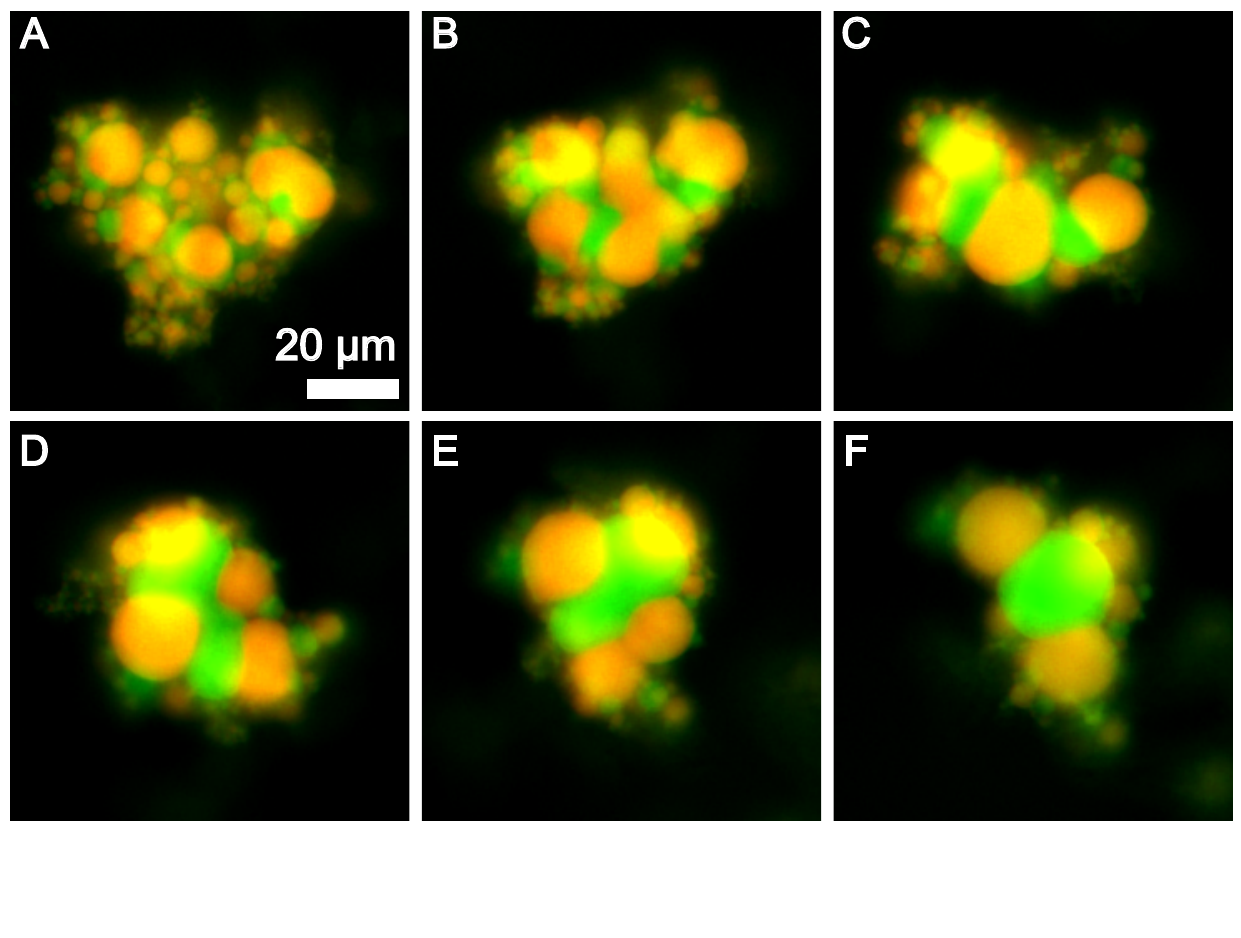}
  \caption{
  Time series of fluorescent images of droplets formed in a mixture, at \SI{1}{M} NaCl, of PLL (\SI{19.2}{\micro M}, green) with orthogonal pNS$_1$ (\SI{5}{\micro M}, red) and pNS$_2$ (\SI{5}{\micro M}, unlabeled), showing multi-phase clusters that coarsen over time. With this labeling scheme, the droplets enriched in PLL+pNS1 appear orange, while the droplets enriched in PLL+pNS2 appear green. Time points are (A)~0 min, (B)~15 min, (C)~30 min, (D)~45 min (E)~60 min, and (F)~165 min after mixing. The $\sim$hours-long aging process can be seen to be due to droplet coarsening by coalescence, whose slow dynamics we attribute to to the combination of the high viscosity of the dense phases, along with the likely small interfacial tension of the boundary between two highly-similar dense liquids that share components.}
  \label{fig:SI-multiphase-coarsening}
\end{figure}

\clearpage

\section{Overview of the Theoretical Model}

\begin{figure}[h]
  \begin{center}\includegraphics[width=0.75\linewidth]{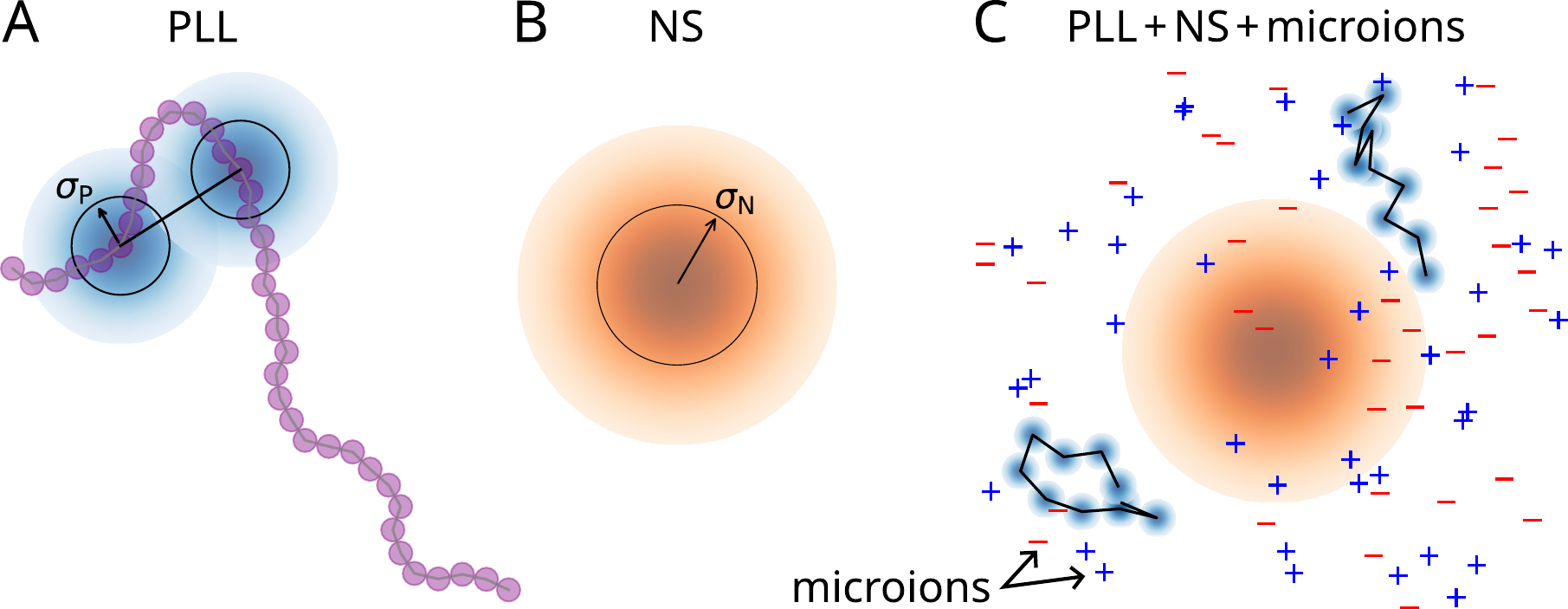}\caption{Schematics of (A)~a PLL polymer (purple monomers) coarse-grained into Gaussian charge blobs with radii $\sigma_{\text{P}}$ (blue blobs), (B)~a DNA nanostar coarse-grained into a Gaussian charge blob with radius $\sigma_{\text{N}}$ (orange blob), and (C)~the full system, comprising coarse-grained PLL polymers, coarse-grained nanostars, and positively and negatively charged point-particle microions.}
    \label{fig:model}
\end{center}
\end{figure}

In our model, NS and PLL polyions are represented at a coarse-grained level by polymer ``blobs''.
Each NS is represented as a single blob, whereas each PLL molecule is represented as a freely jointed chain of $N_\text{p}=10$ Kuhn-length blobs~\cite{shi2013control}.
Each blob carries a Gaussian charge distribution,
\begin{equation}
    q_i(r) = Z_i e \left(\frac{1}{2\pi\sigma_i^2}\right)^\frac{3}{2}\exp{(-r^2/2\sigma_i^2}),
    \label{eq:gaussian_charge}
\end{equation}
where we use the subscript $i$ to indicate NS (N) and PLL (P) species (\figref{fig:model}).
A blob of species $i$ carries a total charge of $Z_i e$ that is spatially distributed according to the blob radius $\sigma_i$.
By contrast, microcations~($+$) and microanions~($-$) are treated as point particles.
We do not distinguish between the different sources of microions, which include both counterions and added monovalent salt, in this model.
The implicit solvent is treated as a continuum with relative permittivity $\epsilon$.
The number densities of these four ion species, $\rho_{\text{N}}$, $\rho_{\text{P}}$, $\rho_{+}$, and $\rho_{-}$, are constrained by the electroneutrality condition
\begin{equation}
  Z_{\rm N} \rho_{\rm N}+\NP Z_{\rm P} \rho_{\rm P}+\rho_+-\rho_- = 0.
  \label{eq:neutrality}
\end{equation}

The full Hamiltonian of the system includes a polyion contribution $H_\text{p}$ that depends only on the polyion blob positions, $\{\bm{R}_i\}$; a microion contribution $H_\text{m}$ that depends only on the microion positions, $\{\bm{r}_i\}$; and a polyion--microion coupling contribution $H_\text{mp}$,
\begin{equation}
  H = H_\text{p}(\{\bm{R}_i\}) + H_\text{m}(\{\bm{r}_i\}) + H_\text{mp}(\{\bm{R}_i\},\{\bm{r}_j\}).
  \label{eq:H_full}
\end{equation}
In what follows, we derive an effective Hamiltonian that depends only on the polyion blob positions,
\begin{equation}
  H_\text{eff}(\{\bm{R}_i\}) = H_\text{p}(\{\bm{R}_i\}) + F_\text{m}(\{\bm{R}_i\}).
  \label{eq:H_CG}
\end{equation}
The polyion Hamiltonian $H_\text{p}(\{\bm{R}_i\})$ consists of several parts, including the kinetic energy, $K_\text{p}$; the excluded volume potential, $U_\text{ev}$; the association potential, $U_\text{assoc}$; and the unscreened Coulombic potential due to polyion--polyion interactions, $U_\text{C}$,
\begin{equation}
  H_\text{p} = K_{\text{p}} + U_{\text{ev}} + U_{\text{assoc}} + U_{\text{C}}.
  \label{eq:H_other}
\end{equation}
The microion-related free energy $F_\text{m}(\{\bm{R}_i\})$ must be obtained by integrating out the microion positions, $\{\bm{r}_i\}$, given fixed polyion positions, $\{\bm{R}_i\}$.

\begin{figure}[h]
\centering
\includegraphics[width=\linewidth]{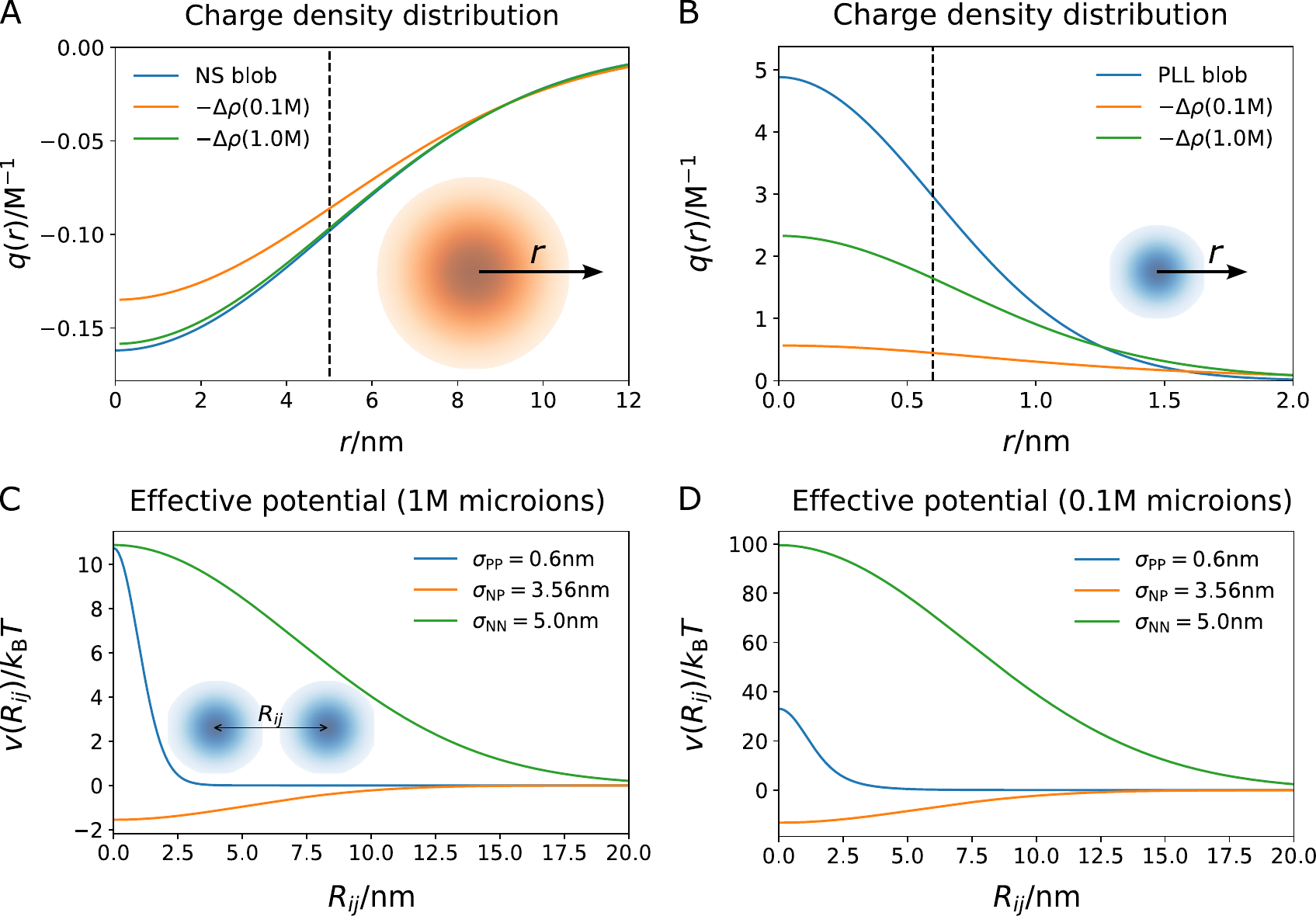}
\caption{Polyion charge distributions, microion charge density distributions, and the resulting screened electrostatic potentials predicted by CDFT.
  (A)~The polyion charge distribution for a NS blob and the predicted charge density distribution, $\Delta \rho(r)\equiv \rho_+(r)-\rho_-(r)$, of the surrounding microions at \SI{0.1}{M} and \SI{1}{M} salt concentrations.  The vertical dashed line represents the NS radius, $\sigma_{\text{N}}$.
  (B)~The polyion charge distribution for a PLL blob and the predicted charge density distribution of the surrounding microions at \SI{0.1}{M} and \SI{1}{M} salt concentrations. The microion distributions are obtained through the Fourier transform of \eqref{eq:EL-solution}.
  The vertical dashed line represents the PLL blob radius, $\sigma_{\text{P}}$.
  (C)~The effective potentials between polyion blobs at \SI{1}{M} salt concentration as a function of the center-to-center distance, $R_{ij}$, given by \eqref{eq:DLVO}.
  (D)~The effective potentials between polyion blobs at \SI{0.1}{M} salt concentration.}
\label{fig:CDFT}
\end{figure}

To proceed, we first evaluate the unscreened polyion Coulombic potential $U_\text{C}$ using the convolution theorem and Parseval’s relation,
\begin{eqnarray}
  U_{\text{C}} &=& \sum_{i<j}^N \int d\bm{s}\int d\bm{s'}\, q_i(\bm{s}-\bm{R_i})q_j(\bm{s'}-\bm{R_j})\frac{1}{\epsilon|\bm{s}-\bm{s'}|}\nonumber \\
  &=& \frac{1}{(2\pi)^3}\sum_{i<j}^N \int d\bm{k}\, q_{i,\bm{k}}q_{j,\bm{k}}\frac{4\pi}{\epsilon k^2} \exp\left[i\bm{k}\cdot(\bm{R_i}-\bm{R_j})\right]\nonumber \\
  &=&\sum_{i<j}^N \frac{Z_iZ_je^2}{\epsilon|\bm{R_i}-\bm{R_j}|}\text{erf}\left(\frac{|\bm{R_i}-\bm{R_j}|}{2\sigma_{ij}}\right),
    \label{eq:bare}
\end{eqnarray}
where $\sigma_{ij}\equiv\sqrt{(\sigma^2_i+\sigma^2_j)/2}$.
We then obtain the microion-related free energy, $F_{\text{m}}$, by minimizing a free energy functional based on Classical Density Functional Theory (CDFT)~\cite{gaussian-blob,hard-sphere-colloidal},
\begin{equation}
  F_\text{m}(\{\bm{R}_i \}) = \min \mathcal{F} [\rho_+ (\bm{r}), \rho_- (\bm{r})],
\end{equation}
where $\rho_+ (\bm{r})$ and $\rho_- (\bm{r})$ are fields representing the concentrations of positive and negative microions, respectively (\figref{fig:CDFT}).
See SI~\secref{SI:derivation} for details of this calculation.
The microion-related free energy includes a state-dependent ``volume term'', $\Phi$, and pairwise screened electrostatic potentials between polyion blobs, $u_{ij}^{\tmop{eff}}$,
\begin{equation}
  F_{\tmop{m}}(\{ \bm{R}_i \}) = \Phi + \sum_{i < j}^N u_{ij}^{\tmop{eff}} (R_{i j}) - U_\text{C},
\label{eq:H_m}
\end{equation}
where $R_{i j} \equiv |\bm{R}_j-\bm{R}_i|$.
The volume term, which does not depend on the positions of the polyions, takes the form
\begin{eqnarray}
  \Phi & = & F_{\tmop{id}} (V, T, \rho_+) + F_{\tmop{id}} (V, T, \rho_-) + F_{\text{ev}}(V,T,\rho_{\text{N}},\rho_{\text{P}},\rho_+,\rho_-)
  \nonumber\\
  &  & - \rho_\text{N} V\frac{Z^2_\text{N} e^2 \kappa}{2 \epsilon} \exp{(\kappa^2\sigma_\text{N}^2)}~\text{erfc}(\kappa \sigma_\text{N}) -\NP\rho_\text{P} V\frac{Z^2_\text{P} e^2 \kappa}{2 \epsilon} \exp{(\kappa^2\sigma_\text{P}^2)}~\text{erfc}(\kappa \sigma_\text{P}) \nonumber\\
  &  & - \frac{V k_\text{B} T}{2 (\rho_++\rho_-)} \left(Z_{\rm N} \rho_{\rm N}+\NP Z_{\rm P} \rho_{\rm P}\right)^2,
  \label{eq:volume}
\end{eqnarray}
where $F_{\text{id}}$ is the ideal free energy of a gas at constant volume $V$, temperature $T$, and mean concentration $\rho_\pm$, and $F_{\text{ev}}$ is the contribution to the free energy due to excluded volume interactions between microions and polyions.
The inverse Debye screening length $\kappa$ is defined as
\begin{equation} \kappa \equiv \left[ 4\pi l_\text{B}(\rho_+ + \rho_-) \right]^{1/2}, \end{equation}
where $l_\text{B}$ is the Bjerrum length, $l_\text{B} \equiv e^2 / \epsilon k_\text{B}T$.
The screened electrostatic potential between the polyion blobs $i$ and $j$ is
\begin{eqnarray}
  u_{ij}^{\tmop{eff}} (R_{ij}) = \frac{Z_i Z_j e^2}{\epsilon R_{ij}} \frac{\exp (
    \kappa^2\sigma^2_{ij})}{2}&\!\![\exp{(-\kappa R_{ij})}~\text{erfc}(\kappa \sigma_{ij}-R_{ij}/2\sigma_{ij}) \nonumber \\
    &\;\;-\exp{(\kappa R_{ij})}~\text{erfc}(\kappa \sigma_{ij}+R_{ij}/2\sigma_{ij})].
  \label{eq:DLVO}
\end{eqnarray}
In these expressions, the complementary error function is defined as
\begin{equation}
  \text{erfc}(z) = \frac{2}{\sqrt{\pi}}\int_z^\infty e^{-t^2}dt.
\end{equation}

At large distances between polyion blobs, the screened electrostatic potential in \eqref{eq:DLVO} has the asymptotic formula
\begin{equation}
    u_{ij}^{\text{eff}}(R_{ij}) \sim \frac{Z_i Z_j e^2}{\epsilon} \,
\exp\left( \kappa^2 \sigma_{ij}^2 \right)\frac{\exp(-\kappa R_{ij})}{R_{ij}}
\label{eq:asym}
\end{equation}
as $R_{ij} \to \infty$.
We notice that \eqref{eq:asym} has the same form as the well known DLVO potential~\cite{derjaguin1941theory,verwey1947theory}, but with a different prefactor.
In fact, the DLVO potential can be derived in a similar way from CDFT, in which case the polyions are represented by hard-sphere colloidal particles with charges uniformly distributed on the surface~\cite{hard-sphere-colloidal}.
The different prefactor reflects the influence of the continuous charge distribution that we assume for each polyion blob.

Finally, we evaluate the free-energy density (i.e., the free energy per unit volume) of the full system, $f$, using the effective Hamiltonian $H_{\text{eff}}(\{\bm{R}_i\})$.
We approximate the contribution to $f$ due to the pairwise polyion--polyion screened electrostatic potential using the mean-field expression
\begin{eqnarray}
    f_\text{mf} &=& 2\pi \sum_{i,j} \rho_i \rho_j \int_0^\infty dR_{ij}\, u_{ij}^\text{eff}(R_{ij}) R_{ij}^2 \nonumber\\ &=& \frac{k_\text{B} T}{2 (\rho_++\rho_-)} \left(Z_{\rm N} \rho_{\rm N}+\NP Z_{\rm P} \rho_{\rm P}\right)^2.
\end{eqnarray}
This expression exactly cancels the last contribution to the volume term, \eqref{eq:volume}.
Moreover, because the unscreened Coulombic potential in $H_{\text{p}}$, \eqref{eq:H_other}, is canceled by $F_{\text{m}}$, \eqref{eq:H_m}, the only surviving contributions due to electrostatics are contained in the ``self-energy'' terms in \eqref{eq:volume},
\begin{equation}
  f_\text{self} = -\frac{k_{\text{B}} T \kappa l_\text{B}}{2}\left[\rhoN Z^2_\text{N}  \exp{(\kappa^2\sigma_\text{N}^2)}~\text{erfc}(\kappa \sigma_\text{N})+\NP\rhoP Z^2_\text{P}  \exp{(\kappa^2\sigma_\text{P}^2)}~\text{erfc}(\kappa \sigma_\text{P})\right].
\end{equation}
The other, non-electrostatic contributions to $f$ arise from the remaining terms in \eqref{eq:H_other} and \eqref{eq:volume}.
Taken together, we write the complete mean-field free-energy density of the system as
\begin{equation}
  f = f_\text{poly}+f_\text{micro}+f_\text{self}+f_\text{assoc},
  \label{eq:free_energy}
\end{equation}
where $f_{\text{poly}}$ accounts for the polyion ideal free energy and the polyion excluded volume potential $U_{\text{ev}}$ in \eqref{eq:H_other}, $f_{\text{micro}}$ accounts for the microion ideal free energy and the excluded volume interactions between microions and polyions in \eqref{eq:volume}, and $f_{\text{assoc}}$ accounts for the association potential between NS sticky ends in \eqref{eq:H_other}.
Expressions for these terms, which are evaluated using a previously developed extension of the Flory--Huggins theory~\cite{li2023interplay}, the Widom insertion method~\cite{widom1963some}, and Wertheim's thermodynamic perturbation theory~\cite{wertheim1984fluids}, respectively, are provided in the Methods section of the main text.

\section{Parametrization of the Theoretical Model}

We summarize the model parameters and provide explanations for their values in Table~\ref{tab:parameters}.

\begin{table}[h]
\centering
  \begin{tabular}{|c|c|c|}
    \hline
    Parameter & Value & Explanation\\
    \hline
    $\sigma_\text{N}$ & \SI{5}{nm} & Approximate hydrodynamic radius ($\sim$\SI{4.7}{nm}~\cite{biffi2013phase})\\
    \hline
    $\sigma_\text{P}$ & \SI{0.6}{nm} & Comparable to the persistence length ($\sim$\SI{1}{nm}~\cite{shi2013control})\\
    \hline
    $N_\text{P}$ & 10 & The number of Kuhn segments assuming $\sim$10 lysine monomers per Kuhn segment~\cite{shi2013control}\\
    \hline
    $Z_\text{N}$ & $-192$ & The total charge per NS\\
    \hline
    $Z_\text{P}$ & 10 & The total charge per PLL Kuhn segment\\
    \hline
    $l_\text{B}$ & \SI{0.71}{nm} & Bjerrum length for pure water at 25$^{\circ}$C\\
    \hline
    $f_\text{N}$ & 0.01 & NS blobs are mostly solvent, with DNA occupying a relatively small volume fraction.\\
    \hline
    $f_\text{P}$ & 0.5 & Roughly half of the volume of each PLL blob is sterically inaccessible to microions.\\
    \hline
  \end{tabular}
  \caption{The physics-based and system-specific constants used to parametrize the theoretical model.}
  \label{tab:parameters}
\end{table}
\FloatBarrier

To parametrize the association contribution to the mean-field free-energy density, $f_{\text{assoc}}$, we compute the dimensionless association strength due to sticky-end base pairing, $\Delta$.
This parameter is utilized in Wertheim's thermodynamic perturbation theory, as described in the Methods section of the main text.
At a standard \SI{1}{M} monovalent cation concentration in aqueous solution, this parameter is
\begin{equation}
  \Delta_0 = \frac{m\Omega}{4\pi}\exp({-\beta \Delta G^\circ})v_\text{N}^{-1} \cdot\text{M}^{-1},
  \label{eq:delta0}
\end{equation}
where $m = 4$ is the number of sticky ends per pNS, $\Omega$ is the solid angle accessible to a sticky end on a pNS blob, and $\vN = 4\pi\sigma_{\text{N}}^3/3$ is the pervaded volume of a NS.
The sequence and temperature-dependent hybridization free energy $\Delta G^\circ$ is calculated using NUPACK~\cite{NUPACK2020}.
We then apply an empirical salt correction~\cite{santalucia2004thermodynamics}
\begin{equation}
  \Delta = \Delta_0 \left(\rho_+ \cdot \text{M}^{-1}\right)^{\alpha},
\end{equation}
where $\alpha = 1.15$, to obtain the dimensionless association strength $\Delta$ at nonstandard salt concentrations.
This correction accounts for the effective strengthening of DNA hybridization due to screening of the charges on the DNA sticky ends by added salt.

\begin{figure}[!htb]
\centering
\includegraphics[width=0.95\linewidth]{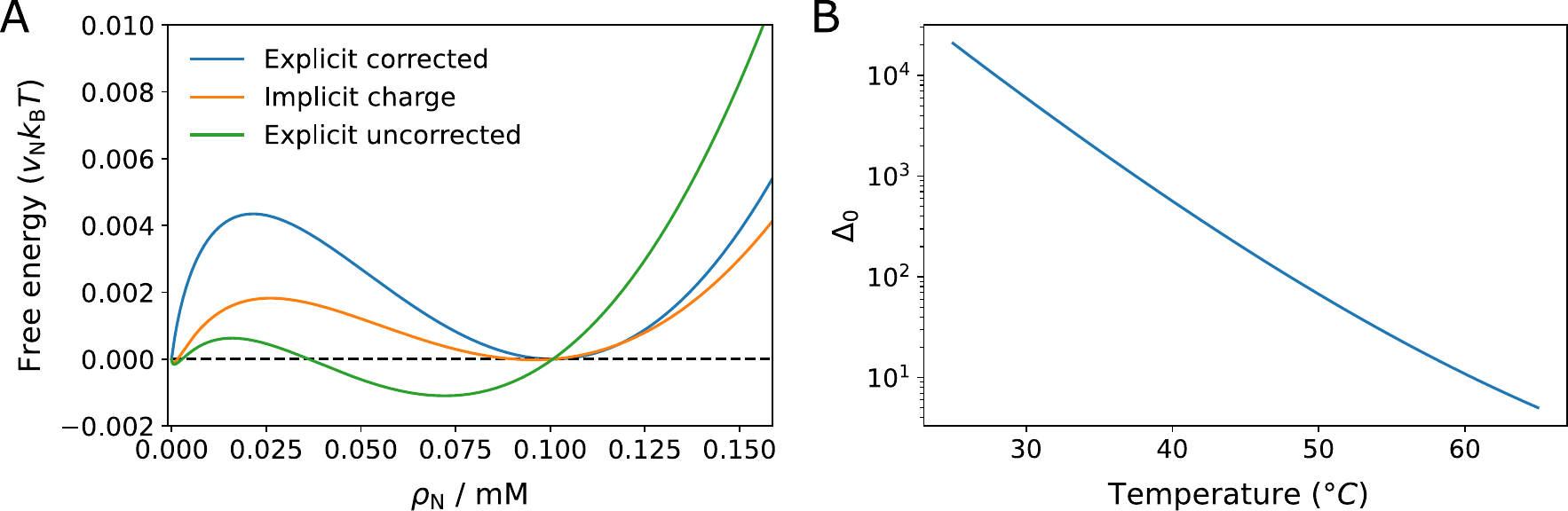}
\caption{Parameterization of the dimensionless association strength due to sticky-end base pairing at \SI{1}{M} NaCl, $\Delta_0$.
  (A)~Comparison of predicted pNS-only free-energy landscapes with explicit charges (i.e., with the physical value of $\ZN$ given in Table~\ref{tab:parameters}) and implicit charges on the sticky ends (i.e., with $\ZN = -192+4\times7=-164$) at \SI{25}{\celsius}.  Explicit-charge landscapes are shown both with the corrected $\Delta G^\circ$ value (blue curve) and with the NUPACK-predicted $\Delta G^\circ$ value (green curve).  The implicit-charge landscape is computed with the NUPACK-predicted $\Delta G^\circ$ value (orange curve).  The correction to $\Delta G^\circ$ is chosen such that the explicit-charge and implicit-charge landscapes predict the same coexistence pNS concentrations; the blue and orange curves have been shifted by a linear term per the common-tangent construction, so that the coexisting phases coincide with local minima on these curves.
  (B)~$\Delta_0$ as a function of temperature.}
\label{fig:delta0}
\end{figure}

We note that the standard hybridization free energy, $\Delta G^\circ$, already includes the effects of screened electrostatics involving the DNA sticky ends at \SI{1}{M} salt.
However, our theory also considers these charges in the electrostatic contributions to our mean-field theory, since the sticky-end charges are included in $Z_{\text{N}}$.
We therefore strengthen the standard hybridization free energy relative to the NUPACK prediction to avoid double counting.
By comparing the predicted free-energy landscapes of a pNS-only solution with and without explicit charges on the sticky ends (i.e., with $\ZN=-192$, and with $\ZN = -192+4\times7=-164$, respectively) at \SI{1}{M} salt, we determine that the NUPACK-predicted value of $\Delta G^\circ$ should be decreased by $3.06 k_{\text{B}}T$ in the presence of explicit charges (\figref{fig:delta0}A).
Finally, we estimate that the solid angle accessible to an individual sticky end on a pNS is approximately $\Omega = \pi / 4$ based on our previous work~\cite{hegde2024competition}.
Applying this parametrization approach to the experimental palindromic sticky-end sequence (including the first non-hybridizing base), 5'-{\fontfamily{qcr}\selectfont {\textit{CGATCGA}}}-3', results in the temperature-dependent $\Delta_0$ values shown in \figref{fig:delta0}B.

\section{Phase-coexistence and minimum-free-energy-path calculations}

Using the free-energy density $f$, we calculate the conditions for phase coexistence following the numerical strategy described in our previous work~\cite{jacobs2023theory}.
We reproduce the details of this numerical approach here for completeness.
Specifically, we solve for the molecular number densities, $\{\vec{\rho}^{(k)}\}$, and the mole fractions of the coexisting phases, $\{x^{(k)}\}$, given the total molecular number densities, $\vec{\rho}_{\tmop{tot}}$.
Mass conservation requires that
\begin{equation}
  \vec{\rho}_{\text{tot}} = \sum^K_{k = 1} x^{(k)} \vec{\rho}^{(k)},
\label{eq:mass-conserv}
\end{equation}
if there are $K$ phases in coexistence.

We now consider the conditions for phase equilibrium in a mixture with $N=3$ independent components (NS, PLL, and microcations), since the microanion concentration is constrained by the charge neutrality condition in \eqref{eq:neutrality}.
The grand-potential density is related to the free-energy density via ${\omega = f - \sum_{i=1}^N (\partial f / \partial \rho_i) \rho_i}$, where the chemical potentials of the non-solvent molecular species are ${\mu_i =  \partial f / \partial \rho_i}$.
Coexisting phases are located at minima of $\omega$, ensuring that all components have equal chemical potentials in all phases.
Phase equilibrium also requires equal pressures across all coexisting phases, implying that the $\omega(\vec\rho)$ has the same value at all local minima.
Together, these conditions require
\begin{equation}
  \omega(\vec{\rho}^{(k)}) = \min \omega(\vec{\rho}; \vec{\mu}).
  \label{eq:phase-equilibrium}
\end{equation}
We solve \eqref{eq:phase-equilibrium} numerically by minimizing the norm of the residual errors of \eqref{eq:mass-conserv} and \eqref{eq:phase-equilibrium} iteratively.
At each iteration, we first locate the local minima of the grand potential, $\vec{\rho}^{(k)} = \arg \min_{\vec{\rho}} \omega(\vec{\rho};\vec{\mu})$ for all phases $k=1,\ldots,K$, using the Nelder--Mead method~\cite{nelder-mead}.
We then update $\vec{\mu}$ and $\{x^{(k)}\}$ using the modified Powell method~\cite{hybrmethod}.

Success of this optimization procedure requires that the initial estimates of $\{\vec\rho^{(k)}\}$ are not too far from the values at coexistence.
We obtain an initial guess for $\{\vec\rho^{(k)}\}$ from the convex hull method~\cite{mao2019phase,jacobs2023theory}, in which we compute the convex hull of points on a discretized $(N+1)$-dimensional free energy surface.
We initialize our optimization procedure with $N + 1$ vectors $\{\vec{\rho}^{(k)}\}$ that correspond to the vertices of the convex hull facet that encompasses $\vec\rho_{\text{tot}}$.
From the linear equation that defines this facet, we also obtain initial guesses for $\vec{\mu}$ and $\{x^{(k)}\}$.
When the number of coexisting phases $K$ is less than $N+1$, some of the $N+1$ vectors $\{\vec{\rho}^{(k)}\}$ are identical to within numerical tolerance after optimization; in this case, we restart the optimization procedure with a unique set of vectors and the corresponding $\vec{\mu}$ and $\{x^{(k)}\}$.
In this way, we determine the number of coexisting phases, $K$, as well as the molecular number-density vectors, $\{\vec\rho^{(k)}\}$, and chemical potentials, $\vec\mu$, at phase coexistence.

To investigate the system's phase-separation kinetics, we examine the grand potential at fixed chemical potentials corresponding to a homogeneous system with the total number densities $\vec\rho_{\text{tot}}$,
\begin{equation}
    \omega_0(\vec{\rho}) = f - \sum_{i=1}^N (\partial f / \partial \rho_i)_{\vec{\rho}_{\text{tot}}} \rho_i.
\end{equation}
We then compute the minimum-free-energy path (MFEP) to each of the condensed phases using the zero-temperature string method~\cite{string2007}.
Following the terminology in Ref.~\cite{string2007}, we use a total of $M=100$ points on a string between two local minima on the free energy surface.
In the ``evolution step'' of the algorithm, the points evolve according to the gradient descent method:
\begin{equation}
  \vec{\rho}^{\;(l + 1)}_k =\vec{\rho}^{\;(l)}_k - \Delta t\left(
  \frac{\partial \omega_0}{\partial \vec{\rho}} \right)_{\!\vec\rho_k^{\,(l)}}\!,
\label{eq:GD}
\end{equation}
where the step size takes $\Delta t=5\times10^{-4}$.
Then in the ``reparametrization step'', we reparametrize the string such that the $M$ points are equally spaced with respect to arc length along the string.
We check for convergence by measuring the norm of the displacement of all points from their positions in the previous iteration, and we use a convergence tolerance of $\text{TOL}=5\times10^{-6}$.

\section{Dominance analysis} \label{sec:SI-dominance}
To quantify the contribution of each species to the driving force for phase separation in the theoretical model, we perform a ``dominance'' analysis following Ref.~\cite{qian2024dominance}.
This analysis decomposes the total free-energy change due to phase separation into contributions from the three independent species, N, P, and microcation (+).
The total free-energy change due to phase separation is $\Delta f\equiv (1-v)f(\vec\rho^{\,\text{(dilute)}})+vf(\vec\rho^{\,\text{(condensed)}})-f(\vec\rho^{\,\text{(parent)}})$, where $v$ is the volume fraction of the condensed phase.
The contributions from the species $\alpha \in \{\text{N}, \text{P}, +\}$, denoted by $\Delta f^\alpha$, sum to the total free-energy change, such that $\sum_\alpha \Delta f^\alpha = \Delta f$.
Because these free-energy changes depend on where the parent concentration lies within the coexistence region in a multicomponent mixture, Ref.~\cite{qian2024dominance} defines a unique ``dominance metric'', $D^\alpha$, at the point where the tie line intersects the dilute branch of the binodal, resulting in an infinitesimal volume fraction of the coexisting condensed phase.
The dominance metric is thus defined as
\begin{equation}
D^{\alpha} \equiv \lim_{v \to 0} \frac{\Delta f^{\alpha}}{\Delta f}
= \frac{\sum_{\beta \in \{\text{N},\text{P},+\}} \Delta\rho_\alpha \Delta\rho_\beta \left.(\partial_{\alpha}\partial_{\beta} f)\right|_{\vec\rho^{\;\text{(dilute)}}}}{\sum_{\beta \in \{\text{N},\text{P},+\}} \sum_{\gamma \in \{\text{N},\text{P},+\}} \Delta\rho_\beta \Delta\rho_\gamma \left.(\partial_{\beta}\partial_{\gamma} f)\right|_{\vec\rho^{\;\text{(dilute)}}}} \,,
\label{eq:dominance}
\end{equation}
where $\Delta\vec\rho \equiv \vec\rho^{\,\text{(condensed)}} - \vec\rho^{\,\text{(dilute)}}$ is the tie-line vector between the dilute phase and the condensed phase.
We use \eqref{eq:dominance} to evaluate $D^\alpha$ directly from the Hessian matrix of the free-energy density given by \eqref{eq:free_energy}.
  
  The results of this analysis are shown for the macromolecularly charge-balanced parent composition, $\rhoN = \SI{10}{\micro M}$ and $\rhoP = \SI{19.2}{\micro M}$, at two temperatures in Fig.~\ref{fig:SI-dominance}.
  Dominance metrics are only plotted for the two macromolecular species N and P, since we find that $|D^+| < 8\times10^{-3}$ for these conditions.
  At $T = \SI{25}{\celsius}$ (Fig.~\ref{fig:SI-dominance}A), two two-phase regions are predicted at low and high salt concentrations, whereas a three-phase coexistence region is predicted at intermediate salt concentrations (cf.~Fig.~1I in the main text).
  At low salt concentrations, the dominance analysis shows that the pNS+PLL coacervate is stabilized by both pNS and PLL ($D^{\text{N}} > 0$ and $D^{\text{P}} > 0$), as expected for coacervation driven by heterotypic electrostatic interactions.
  However, the relative contribution from pNS increases with increasing salt concentration.
  This result is consistent with an increasing role of homotypic base pairing interactions with increasing salt concentration.
  By contrast, the pNS-dominated condensate at high salt concentrations is overwhelmingly stabilized by pNS, even though PLL has a partition coefficient greater than one under these conditions (see Fig.~\ref{fig:SI-pc}) and thus also partitions into pNS-dominated condensates at equilibrium.

  In the three-phase region at intermediate salt concentrations at $T = \SI{25}{\celsius}$, we compute the dominance metric separately for each condensed phase.
  We find that the trends are consistent with the adjacent two-phase regions.
  Notably, pNS provides the greater driving force for phase separation for both condensed phases in this region.
  This result indicates that even though PLL partitions into the pNS+PLL coacervate to a much greater extent than it does the pNS-dominated condensate at these conditions, homotypic base pairing interactions play an essential role in stabilizing both condensed phases when three-phase coexistence is observed.
  
  At $T = \SI{55}{\celsius}$ (Fig.~\ref{fig:SI-dominance}B), only the pNS+PLL coacervate is stable at low-to-intermediate salt concentrations.
  Consistent with the analysis at $T = \SI{25}{\celsius}$, we find that both pNS and PLL contribute to the driving force for phase separation for salt concentrations at and above $\rho_{\text{U}}$, which is consistent with coacervation due in part to heterotypic electrostatic interactions.
  This observation is notable since the heterotypic electrostatic interactions alone are insufficient to drive phase separation at these salt concentrations (cf.~Fig.~1H in the main text).
  Moreover, the temperature response of $D^{\text{N}}$ is weaker at $T = \SI{55}{\celsius}$ than at $T = \SI{25}{\celsius}$ when $\rhoS \gtrsim \rho_{\text{U}}$, indicating a decreasing contribution from homotypic base pairing interactions upon increasing the temperature.
  This observation is consistent with the weakening of homotypic base pairing interactions with increasing temperature, which are insufficient to drive phase separation in the absence of heterotypic electrostatic interactions at $T = \SI{55}{\celsius}$ (cf.~Fig.~1H in the main text).

\section{Derivation of the Microion-related Free Energy Using CDFT}
\label{SI:derivation}

In this section, we derive the microion-related free energy, $F_{\text{m}}(\{\bm{R}_i\})$, by extending the classical density functional theory (CDFT) derivation in Ref.~\cite{gaussian-blob} to multicomponent polyion systems.
We write the free energy functional as
\begin{equation}
  \mathcal{F} [\rho_+ (\bm{r}), \rho_- (\bm{r})]
  =\mathcal{F}_{\tmop{id}} +\mathcal{F}_{\tmop{m}}
  +\mathcal{F}_{\tmop{pm}},
\end{equation}
where $\rho_{\pm} (\bm{r}) \equiv \rho^{(1)}_{\pm} (\bm{r})$ is the number density~\cite{HANSEN2013203} of microions with charge $+1$ or $-1$, respectively, and we neglect microion correlations.
We consider local microion inhomogeneities,
\begin{equation}
  \Delta \rho_{\pm} (\bm{r}) \equiv
  \rho_{\pm} (\bm{r}) - \rho_{\pm},
\end{equation}
where $\rho_{\pm} = V^{-1} \int d\bm{r}\, \rho_{\pm} (\bm{r})$ is the mean number density of each microion species.
This allows us to expand the ideal gas part of the free-energy functional to quadratic order,
\begin{equation}
  \mathcal{F}_{\tmop{id}} = \sum_{\alpha = \pm} \left[ F_{\tmop{id}} (V, T,
  \rho_{\alpha}) + \frac{\kB T}{2 \rho_{\alpha}} \int d\bm{s}\, (\Delta
  \rho_{\alpha} (\bm{s}))^2 \right],
\end{equation}
where $F_{\tmop{id}} (V, T, \rho_{\alpha}) = V \kB T \rho_{\alpha} [\ln (\rho_{\alpha}
\Lambda^3_{\alpha}) - 1]$ is the homogeneous ideal gas contribution from
microion species $\alpha = \pm$.
The microion--microion Coulomb interactions contribute to the free-energy functional in the Hartree form,
\begin{equation}
  \mathcal{F}_{\tmop{m}} = \frac{e^2}{2 \epsilon} \int d\tmmathbf{s} \int
  d\bm{s}'\, \frac{[\rho_+ (\bm{s}) - \rho_- (\bm{s})] [\rho_+
  (\bm{s}') - \rho_- (\bm{s}')]}{| \bm{s}-\bm{s}' |}.
\end{equation}
Meanwhile, the microion--polyion contribution takes the form
\begin{equation}
  \mathcal{F}_{\tmop{pm}} = \sum_{\alpha = \pm} \int d\bm{s}\,
  \rho_{\alpha} (\bm{s}) U_{\alpha} (\bm{s})- (\rho_++\rho_-)\ln{(1 - \eta)},
\end{equation}
where the potential experienced by the microions due to the $N$ polyions with Gaussian charge distributions, \eqref{eq:gaussian_charge}, is
\begin{equation}
    U_+(\bm{r}) = -U_-(\bm{r}) = \sum_{i=1}^N Z_i \frac{e^2}{\epsilon| \bm{r}-\bm{R}_i |}\text{erf}\left(\frac{| \bm{r}-\bm{R}_i |}{\sqrt{2}\sigma_i}\right).
    \label{eq:potential}
\end{equation}
The final term in the expression for $\mathcal{F}_{\text{pm}}$ describes the polyion--microion excluded volume contribution to the free energy, where $\eta$ is the volume fraction of the system that is sterically inaccessible to microions.
As discussed in the Methods section of the main text, this term is derived using the Widom insertion method~\cite{widom1963some} assuming that the volume occupied by each microion is negligible.

We diagonalize $\mathcal{F}$ by defining
\begin{equation}
  \rho (\bm{r}) \equiv \rho_+ (\bm{r}) - \rho_- (\bm{r})
\end{equation}
and
\begin{equation}
  \delta (\bm{r}) \equiv \frac{\rho_- \rho_+ (\bm{r}) + \rho_+ \rho_-
    (\bm{r})}{\rho_+ + \rho_-}.
\end{equation}
The functional can then be written as
\begin{equation}
  \mathcal{F} [\rho, \delta] = F_{\tmop{id}} (V, T, \rho_+) + F_{\tmop{id}} (V,
  T, \rho_-) - (\rho_++\rho_-)\ln{(1 - \eta)}+\mathcal{F}_{\tmop{1}} [\rho] +\mathcal{F}_{\tmop{2}} [\delta] ,
  \label{eq:F_separation}
\end{equation}
where
\begin{equation}
  \mathcal{F}_{\tmop{1}} [\rho] = \frac{\kB T}{2 (\rho_+ + \rho_-)} \int
  d\bm{s}\, [\rho (\bm{s}) - \bar{\rho}]^2 + \int d\bm{s}\, \rho
  (\bm{s}) U (\bm{s}) + \frac{e^2}{2 \epsilon} \int d\bm{s} \int
  d\bm{s}' \frac{\rho (\bm{s}) \rho (\bm{s}')}{|
  \bm{s}-\bm{s}' |},
\end{equation}
\begin{equation}
  \mathcal{F}_{\tmop{2}} [\delta] = \frac{\kB T}{2} \left( \frac{1}{\rho_+} +
  \frac{1}{\rho_-} \right) \int d\bm{s}\, [\delta (\bm{s}) -
  \bar{\delta}]^2 + \int d\bm{s}\, \delta (\bm{s}) W (\bm{s}),
\end{equation}
and $\bar{\rho}$ and $\bar{\delta}$ are the spatial averages of the variational fields, ${\bar{\rho} \equiv \rho_+ - \rho_-}$ and ${\bar{\delta} \equiv 2 \rho_+ \rho_- / (\rho_++\rho_-)}$.
The definitions of the muticentered functions $U(\bm{r})$ and $W (\bm{r})$ follow from the linear combinations of the external fields,
\begin{equation}
  U (\bm{r}) = \frac{\rho_+ U_+ (\bm{r}) - \rho_- U_- (\bm{r})}{\rho_+ + \rho_-}
\end{equation}
and
\begin{equation}
  W (\bm{r}) = U_+ (\bm{r}) + U_- (\bm{r}).
\end{equation}
Due to the $\pm$ symmetry of \eqref{eq:potential}, we find that $U(\bm{r})=U_+(\bm{r})$ and $W(\bm{r})=0$.
The functional $\mathcal{F}_2[\delta]$ is therefore minimized by setting $\delta(\bm{r})=\bar\delta$, indicating that this contribution to $\mathcal{F}$ vanishes at equilibrium.

To minimize $\mathcal{F}_{\tmop{1}} [\rho]$, we solve the Euler--Lagrange (EL) equation
\begin{equation}
  \mu_{\rho} = \frac{\partial \mathcal{F}_{\tmop{1}} [\rho]}{\partial \rho (\bm{r})}
  =  \frac{\kB T}{n} \Delta \rho (\bm{r}) + U (\bm{r}) + \frac{e^2}{\epsilon} \int d\bm{s}' \frac{\rho (\bm{s}')}{| \bm{r}-\bm{s}' |},
\end{equation}
where $n \equiv \rho_+ +\rho_-$ and $\Delta\rho(\bm{r}) \equiv \rho(\bm{r})-\bar{\rho}$.
To handle divergences in the volume integrals, we employ a screened Coulomb potential, i.e., $\exp(-\lambda r)r^{-1}$, in place of $r^{-1}$ and take the limit $\lambda \rightarrow 0$ later.
We denote the screened form of $U$ by $U^{(\lambda)}$ in what follows.
Using the Fourier transform
\begin{equation}
  f_{\bm{k}} = \int_V d\bm{s}f (\bm{s}) \exp (i\bm{k} \cdot
  \bm{s}),
\end{equation}
we obtain
\begin{equation}
  \mu_{\rho} (2 \pi)^3 \delta (\bm{k}) = \frac{\kB T}{n} \Delta
  \rho_{\bm{k}} + U^{(\lambda)}_{\bm{k}} + \frac{e^2}{\epsilon}
  \frac{4 \pi}{k^2 + \lambda^2} [(2 \pi)^3 \bar{\rho} \delta (\bm{k}) +
  \Delta \rho_{\bm{k}}].
\end{equation}
Then, using $(2 \pi)^3 \delta (\bm{k}= 0) = V$ and $\Delta \rho_{\bm{k}= 0} = 0$, we can write the $\bm{k}= 0$ component as
\begin{equation}
  \mu_{\rho} V = U^{(\lambda)}_{\bm{k}= 0} + \frac{4 \pi e^2}{\epsilon \lambda^2} \bar{\rho} V.
\end{equation}
The explicit solution to the EL equation now reads
\begin{equation}
  \kB T \Delta \rho_{\bm{k}} = - n \frac{k^2 + \lambda^2}{k^2 +
  \kappa^2_{\lambda}} U^{(\lambda)}_{\bm{k}} + \frac{(2 \pi)^3 \delta
  (\bm{k})}{V} n \frac{\lambda^2}{\kappa^2_{\lambda}}
  U^{(\lambda)}_{\bm{k}= 0}, \label{eq:EL-solution}
\end{equation}
where we have introduced the inverse Debye screening length, $\kappa$, and its modified counterpart $\kappa_\lambda$.
These quantities are given by
\begin{equation}
  \kappa^2 \equiv \frac{4 \pi e^2 (\rho_++\rho_-)}{\epsilon \kB T} = 4 \pi l_{\text {B}} (\rho_++\rho_-)
\end{equation}
and $\kappa^2_{\lambda} \equiv \kappa^2 + \lambda^2$, where $l_{\text {B}}$ is the Bjerrum length.
Example plots of $\Delta \rho(r)$, obtained from the inverse Fourier transform of \eqref{eq:EL-solution} with $\lambda \rightarrow 0$, are shown in \figref{fig:CDFT}A,B.

We now evaluate $\mathcal{F}_1$ at its minimum, using \eqref{eq:EL-solution}, to obtain the equilibrium free energy $F_1$.
In its Fourier representation, $F_1$ is
\begin{equation}
  F_{\tmop{1}} = - \frac{\rho_++\rho_-}{2 \kB T} \frac{1}{(2 \pi)^3} \int d\bm{k}
  \frac{k^2 + \lambda^2}{k^2 + \kappa^2_{\lambda}} U^{(\lambda)}_{-\bm{k}}
  U^{(\lambda)}_{\bm{k}}
  + \frac{\rho_++\rho_-}{2 V \kB T} \frac{\lambda^2}{\kappa^2_{\lambda}}
  (U^{(\lambda)}_{\bm{k}= 0})^2 + \bar{\rho} U^{(\lambda)}_{\bm{k}= 0}
  + \frac{2 \pi e^2}{\epsilon \lambda^2} \bar{\rho}^2 V. 
  \label{eq:F_el}
\end{equation}
The first term in \eqref{eq:F_el} can be evaluated as
\begin{eqnarray}
 &\;& - \frac{\rho_++\rho_-}{2 \kB T} \frac{1}{(2 \pi)^3} \int d\bm{k} \frac{k^2 +
  \lambda^2}{k^2 + \kappa^2_{\lambda}} U^{(\lambda)}_{-\bm{k}}
 U^{(\lambda)}_{\bm{k}} \nonumber \\
 &=& - \frac{\rho_++\rho_-}{2 \kB T} \frac{1}{(2 \pi)^3} \int
  d\bm{k} \frac{k^2 + \lambda^2}{k^2 + \kappa^2_{\lambda}} \sum_{i = 1}^N
  \sum_{j = 1}^N u^{(\lambda)}_{i, -\bm{k}} u^{(\lambda)}_{j, \bm{k}}
  \exp (i\bm{k} \cdot \bm{R}_{i j}) \nonumber \\ &=&-\frac{1}{(2\pi)^3}\frac{2\pi}{\epsilon}\int d\bm{k} \left(\frac{1}{k^2}-\frac{1}{k^2+\kappa^2}\right)\sum_{i = 1}^N
  \sum_{j = 1}^N q_{i, \bm{k}}q_{j, \bm{k}}\exp (i\bm{k} \cdot \bm{R}_{i j}),
  \label{eq:firstline}
\end{eqnarray}
where we have used the (screened) Coulombic potential due to polyion blob $i$, ${u^{(\lambda)}_{i, \bm{k}} = 4\pi q_{i, \bm{k}} / \epsilon(k^2+\lambda^2)}$, in accordance with \eqref{eq:potential}, and $q_{i, \bm{k}}$ is the Fourier transform of the polyion blob charge distribution.
The limit $\lambda \to 0$ is taken in the final line.

We first evaluate the pairwise part of \eqref{eq:firstline} for $i \ne j$.
The $1/k^2$ terms in \eqref{eq:firstline} sum to $-U_{\text{C}}$ and thus cancel the unscreened Coulombic potentials in \eqref{eq:bare}.
The remaining terms constitute the effective pair potentials between polyion blobs, $\sum_{i\ne j}u_{ij}^{\text{eff}}(R_{ij})$, where
\begin{eqnarray}
  \label{eq:pair-potential}
  u_{ij}^{\text{eff}}(R_{ij}) &=& \frac{1}{(2\pi)^3}\frac{4\pi}{\epsilon}\int d\bm{k} \frac{1}{k^2+\kappa^2} q_{i, \bm{k}}q_{j, \bm{k}}\exp (i\bm{k} \cdot \bm{R}_{ij})\\&=& \frac{Z_i Z_j e^2 }{\epsilon R_{ij}} \frac{\exp (\kappa^2\sigma^2_{ij})}{2}[\exp{(-\kappa R_{ij})}~\text{erfc}(\kappa \sigma_{ij}-R_{ij}/2\sigma_{ij})-\exp{(\kappa R_{ij})}~\text{erfc}(\kappa \sigma_{ij}+R_{ij}/2\sigma_{ij})]\nonumber.
\end{eqnarray}
We then evaluate the $i=j$ terms in \eqref{eq:firstline}, which correspond to the ``self energies'' due to the interactions between each polyion and its microion cloud.
The total self energy for all $N$ polyions is
\begin{equation}
F_{\text{self}}  = - \rho_\text{N} V\frac{Z^2_\text{N} e^2 \kappa}{2 \epsilon} \exp{(\kappa^2\sigma_\text{N}^2)}~\text{erfc}(\kappa \sigma_\text{N}) -\NP\rho_\text{P} V\frac{Z^2_\text{P} e^2 \kappa}{2 \epsilon} \exp{(\kappa^2\sigma_\text{P}^2)}~\text{erfc}(\kappa \sigma_\text{P}).
    \label{eq:self}
\end{equation}

To evaluate the remaining terms in \eqref{eq:F_el}, we expand $U^{(\lambda)}_{\bm{k}= 0}$ with respect to $\lambda$.
We consider each polyion blob $i$ individually,
\begin{eqnarray}
  u^{(\lambda)}_{i,
  \bm{k}= 0}& = &4\pi \frac{Z_i e^2}{\epsilon} \int_0^\infty \text{erf}\left(\frac{s}{\sqrt{2}\sigma_i} \right) e^{-\lambda s} s ds \nonumber \\ &= &4\pi \frac{Z_i e^2}{\epsilon}\left[\left(\frac{1}{\lambda^2}- \sigma_i^2\right)\exp{\left(\frac{\lambda^2\sigma_i^2}{2}\right)}\text{erfc}\left(\frac{\lambda\sigma_i}{\sqrt{2}}\right)+\sqrt{\frac{2}{\pi}}\frac{\sigma_i}{\lambda}\right] \nonumber\\ &= &4\pi \frac{Z_i e^2}{\epsilon}\left[\frac{1}{\lambda^2}-\frac{\sigma_i^2}{2}+\mathcal{O}(\lambda)\right],
\end{eqnarray}
and then perform the sum over all polyions to obtain
\begin{align}
  U^{(\lambda)}_{\bm{k}=0} 
  &= V \left( \rho_\text{P} N_\text{P} u^{(\lambda)}_{\text{P},\, \bm{k}=0} 
          + \rho_\text{N} u^{(\lambda)}_{\text{N},\, \bm{k}=0} \right) \nonumber \\
  &= 4\pi V \frac{e^2}{\epsilon} \left( 
      \frac{-\bar{\rho}}{\lambda^2} 
      - \frac{ \rho_\text{P} N_\text{P} Z_\text{P} \sigma_\text{P}^2 
             + \rho_\text{N} Z_\text{N} \sigma_\text{N}^2 }{2} 
    \right) 
  + \mathcal{O}(\lambda),
\end{align}
where we have used the electroneutrality condition
\begin{equation}
  \bar{\rho} + \sum^m_{i = 1} Z_i n_i = 0
\end{equation}
in the last line.
Similarly, we expand the term in \eqref{eq:F_el} that includes $(U^{(\lambda)}_{\bm{k}= 0})^2$ to obtain
\begin{eqnarray}
    \frac{\rho_++\rho_-}{2 V \kB T} \frac{\lambda^2}{\kappa^2_{\lambda}}
  (U^{(\lambda)}_{\bm{k}= 0})^2 &=& \frac{1}{2V}\frac{\kappa^2\epsilon}{4\pi e^2}\frac{\lambda^2}{\kappa^2}\left(1-\frac{\lambda^2}{\kappa^2}\right)(U^{(\lambda)}_{\bm{k}= 0})^2+ \mathcal{O}(\lambda) \\ &=&2 \pi V\frac{e^2}{\epsilon} \left[\frac{\bar{\rho}^2}{\lambda^2}-
  \frac{\bar{\rho}^2}{\kappa^2}+\bar{\rho}(\rho_\text{P}N_\text{P}Z_
  \text{P}\sigma_\text{P}^2+\rho_\text{N}Z_\text{N}\sigma_\text{N}^2)\right]+\mathcal{O}(\lambda).\nonumber
\end{eqnarray}
The last three terms of \eqref{eq:F_el} therefore simplify to
\begin{equation}
  \frac{\rho_++\rho_-}{2 V \kB T} \frac{\lambda^2}{\kappa^2_{\lambda}}
  (U^{(\lambda)}_{\bm{k}= 0})^2 + \bar{\rho} U^{(\lambda)}_{\bm{k}= 0}
  + \frac{2 \pi e^2}{\epsilon \lambda^2} \bar{\rho}^2 V = - \frac{V \kB T}{2
  (\rho_++\rho_-)} \bar{\rho}^2.
  \label{eq:secondline}
\end{equation}

Combining the results from \eqref{eq:F_separation}, \eqref{eq:firstline}, \eqref{eq:pair-potential}, \eqref{eq:self}, and \eqref{eq:secondline}, we summarize the minimized free energy functional, which is $F_\text{m}(\{\bm{R}_i\})$ in \eqref{eq:H_CG},
\begin{eqnarray}
      F_\text{m}(\{
  \bm{R}_i \}) &=& \min \mathcal{F} [\rho_+ (\bm{r}), \rho_- (\bm{r})] \nonumber \\&=& F_{\tmop{id}} (V, T, \rho_+) + F_{\tmop{id}} (V, T, \rho_-)\nonumber - (\rho_++\rho_-)\ln{(1 - \eta)}\\&&-U_\text{C}+\sum_{i<j}^N u_{ij}^{\text{eff}}(R_{ij}) + F_{\text{self}} - \frac{V \kB T}{2
  (\rho_++\rho_-)} \bar{\rho}^2.
  \label{eq:H_CG_DFT}
\end{eqnarray}
This expression can be rearranged to separate the $\{\bm{R}_i\}$-dependent terms and the remaining ``volume'' terms, as presented in \eqref{eq:H_m} and \eqref{eq:volume}.

\end{document}